\documentclass[prd,eqsecnum,twocolumn,nofootinbib]{revtex4-2}

\usepackage{amsfonts}
\usepackage{mathtools}
\usepackage{amssymb}
\usepackage{amsthm}
\usepackage{bm}
\usepackage{dcolumn}
\usepackage{epsfig}
\usepackage{graphicx}
\usepackage{graphics}
\usepackage[latin1]{inputenc}
\usepackage{latexsym}
\usepackage{rotating}
\usepackage{xcolor}
\usepackage{gensymb}

\newcommand{\f}[2]{\ensuremath{\frac{#1}{#2}}}
\newcommand{\fpar}[2]{\ensuremath{\left(\frac{#1}{#2}\right)}}

\begin{document}
\title{Transition from adiabatic inspiral to plunge for eccentric binaries}
\author{Devin R.\ Becker}
\affiliation{Department of Physics and MIT Kavli Institute, MIT, Cambridge, MA 02139 USA}
\author{Scott A.\ Hughes}
\affiliation{Department of Physics and MIT Kavli Institute, MIT, Cambridge, MA 02139 USA}
\begin{abstract}
Black hole binaries with small mass ratios will be critical targets for the forthcoming Laser Interferometer Space Antenna (LISA) mission.  They also serve as useful tools for understanding the properties of binaries at general mass ratios.  In its early stages, such a binary's gravitational-wave-driven inspiral can be modeled as the smaller body flowing through a sequence of geodesic orbits of the larger black hole's spacetime.  Its motion through this sequence is determined by the rate at which backreaction changes an orbit's integrals of motion $E$, $L_z$, and $Q$.  Key to the motion being close to a geodesic at any moment is the idea that the effect of backreaction is small compared to a ``restoring force'' arising from the potential which governs geodesic motion. This restoring force holds the small body on a geodesic trajectory as the backreaction causes that geodesic to slowly evolve.  As the inspiraling body approaches the last stable orbit (LSO), the restoring force becomes weaker and the backreaction becomes stronger.  Once the small body evolves past the LSO, its trajectory converges to a plunging geodesic. This work aims to smoothly connect these two disparate regimes: the slowly evolving adiabatic inspiral and the final plunge.  Past work has focused on this transition to plunge for circular systems.  Here, we study the transition for binaries with eccentricity.  A well-defined eccentric transition will make it possible to develop small-mass-ratio binary waveform models that terminate in a physically reasonable way, rather than abruptly terminating as an inspiral-only model ends.  A model that can explore the parameter space of eccentricity may also be useful for understanding the final cycles of eccentric binaries at less extreme mass ratios, such as those likely to be observed by ground-based detectors.
\end{abstract}
\maketitle

\section{Introduction}
\label{sec:intro}

\subsection{Background, motivation, and past work}

The small-mass-ratio limit of the relativistic two-body problem has proven to be a productive laboratory for exploring the properties of binary systems in general relativity. Various aspects of this problem can be solved precisely when one member of the binary is much more massive than the other, especially if the larger member is taken to be a black hole.  The spacetime of such binaries can be regarded as the spacetime of the large black hole, for which we use the exact Kerr spacetime of general relativity \cite{Kerr1963}, plus a contribution from the smaller orbiting body. Due to the small mass ratio, the orbiting body's contribution can be effectively modeled using techniques from black hole perturbation theory.  The importance of small-mass-ratio binaries as sources of gravitational waves (GWs) for future low-frequency GW detectors \cite{LISA2017, Babak2017}, plus the insight they provide into understanding binary systems more generally \cite{Akcay2015, Wardell2023}, has led to excellent progress in this problem in recent years \cite{Pound2022}.

A key feature of the small-mass-ratio limit is that it allows us to clearly distinguish different dynamical regimes in the binary's evolution, in particular an adiabatic \textit{inspiral} and a final \textit{plunge}.  The binary executes an adiabatic inspiral when its members are widely separated.  In this regime, the backreaction of GWs changes the binary's properties slowly.  The system's  evolution can thus be accurately approximated as a smaller body that ``flows'' through a sequence of geodesic orbits, with the rate of flow governed by the backreaction of GW emission \cite{Hughesetal2005}. This sequence of orbits is built using the {\it self force} arising from the secondary's interaction with its own contribution to the spacetime.  These self force corrections include both dissipative contributions which drive the inspiral, as well as conservative contributions that change the nature of the orbit even when radiative backreaction is neglected (see, e.g., Refs.\ \cite{Warburton2012, Osburn2016} for detailed examination of the effect of conservative contributions on Schwarzschild inspiral). In the binary's final moments, it enters the plunge, during which the motion is well approximated by a geodesic which carries the smaller body into the larger black hole's event horizon.

In this work, we seek to understand when adiabatic inspiral ends, and how to smoothly connect inspiral to the final plunge.  Past work \cite{OriThorne2000, ApteHughes2019, CompereKuchler2021} examined this question in the context of quasi-circular inspiral; we summarize the key points of these studies in Appendix \ref{app:circtrans}.  The central idea in these analyses is to compare the rate at which GW-driven backreaction changes the system's properties to the time it takes for a system to settle into a new equilibrium.  Ori and Thorne \cite{OriThorne2000} (hereafter OT00) developed an equation of motion describing the transition regime by expanding the radial geodesic equation near the innermost stable circular orbit (ISCO).  Apte and Hughes \cite{ApteHughes2019} (hereafter AH19) refined the OT00 analysis, and showed that, with slight modifications, it describes all circular orbits, including inclined ones.

Both OT00 and AH19 develop an acceleration equation for the smaller body's radial motion which has the form
\begin{equation}
    \frac{d^2r}{d\lambda^2} = \frac{1}{2}\frac{\partial R}{\partial r}\;.
    \label{eq:radialaccelintro}
\end{equation}
Here, $\lambda$ is a time parameter (``Mino time'') that is especially useful in describing strong-field black hole orbits; the function $R$ depends on the orbital radius $r$ and on the integrals of motion $E$, $L_z$, and $Q$.  Both $\lambda$ and $R$ are defined precisely and discussed in detail in Sec.\ \ref{sec:geods}.

When the integrals of motion are constant, Eq.\ (\ref{eq:radialaccelintro}) describes radial geodesic motion, which is particularly simple for circular orbits --- $r$ is a constant in that limit.  OT00 assumed that $E$ and $L_z$ evolve with a simple, linear time dependence during the (very short) transition\footnote{The OT00 analysis ignores the role of the Carter constant since they only focus on equatorial orbits, for which $Q = 0$.}.  In this circumstance, OT00 showed that there is a universal solution to their transition equation which holds for all spins of the larger black hole, and for all mass ratios small enough for perturbation theory to describe the binary.  AH19 found that, with slight adjustment to the definitions of certain parameters, this universal solution describes the transition even for inclined circular orbits.  More recent work \cite{kuchler2024} suggests that this solution can be considered the leading term in an expansion in a ratio of timescales, and shows how to improve the transition description by iterating to higher orders.

The inspiral, transition, and plunge worldlines generated with the approach of AH19 were used to examine how different ringdown modes are excited as a function of a binary's orbital geometry \cite{Lim2019, Hughes2019}, finding a dependence that has been largely validated by later work analyzing this problem at less extreme mass ratios \cite{Ma2021, Zhu2023}. These worldlines have also proven useful in developing surrogate model waveforms which describe comparable mass systems by applying an extrapolation to the phase and amplitude of waveforms developed in the small-mass-ratio limit \cite{Rifat2020, Islam2022}.

\subsection{This paper: Transition for eccentric inspiral}

Our goal now is to incorporate eccentricity into the transition framework\footnote{A first attempt to model the eccentric transition was presented by Sundararajan in Ref.\ \cite{Sundararajan2008b}.  However, flaws identified in AH19 for Sundararajan's model of the inclined quasi-circular transition are even more prominent for eccentric configurations.}.  Ignoring the backreaction of GWs for a moment, an eccentric orbit does not sit at a fixed radius, but instead oscillates between radii $r_{\rm a}$ (``apoapsis'') and $r_{\rm p}$ (``periapsis'').  Equation (\ref{eq:radialaccelintro}) describes this behavior when the orbital integrals are constant.

Now consider how this changes when radiation reaction is taken into account.  The leading change is a slow evolution to the orbit's apoapsis and periapsis due to the backreaction of GW emission.  For most of the system's evolution, the backreaction-driven acceleration of these turning points is so small that its impact on orbits can be neglected; one can simply treat inspiral as a sequence of geodesic orbits.  As the last stable orbit or LSO (the generalization of the ISCO for eccentric orbits) is approached, this is no longer the case.  When the acceleration of the periapsis due to radiative backreaction is no longer negligible, then the eccentric inspiral has entered the transition from inspiral to plunge.  Just as with quasi-circular inspiral, we need different techniques to describe the binary's evolution through this regime.

In this paper, we examine how and when an eccentric binary enters the transition from inspiral to plunge, and we develop tools to describe this regime.  As with the transition for circular orbits, the foundation of our approach continues to be Eq.\ (\ref{eq:radialaccelintro}).  In the absence of radiation reaction, this equation exactly describes the oscillatory radial motion of a geodesic orbit.  By allowing the orbit's integrals of motion to evolve, just as is done for the circular transition, we show that this equation describes the transition from inspiral to plunge for binaries with eccentricity as well.

An interesting and important result of this analysis is that the eccentric transition does {\it not} share the universal radial behavior found for quasi-circular inspiral.  Consider two binaries that are identical in every way but initial conditions.  These orbits share values of the semi-latus rectum $p$ and eccentricity $e$, and are described by the same geodesic.  However, at $t = 0$, one orbit is close to periapsis and the other is close to apoapsis.  The adiabatic evolution of these binaries is identical --- at adiabatic order, each inspiral evolves through the same sequences $p(t)$ and $e(t)$.  The orbits remain out of phase with one another during inspiral: at any moment, if one is close to apoapsis, then the other is close to periapsis.  As they approach the LSO, the behavior of these binaries diverges significantly: one enters the transition close to periapsis, quickly proceeding to plunge; the other enters the transition close to apoapsis, and must complete another half cycle of radial motion before it can plunge.  The late-time worldlines from transition and plunge of eccentric systems --- and thus the last GWs that these systems produce --- will depend strongly on initial phases describing the system.  One of our goals in this paper is to explore the consequences of this initial-phase dependence for the smaller body's final plunge.  A follow-on analysis will study its repercussions for the waveforms that the system produces, focusing on the observational implications of this behavior.

\subsection{Organization of this paper}

The remainder of this paper works out the details of the eccentric transition.  We begin in Sec.\ \ref{sec:geods} with a brief review of the most essential properties, for the purpose of our analysis, of geodesic orbits in Kerr spacetime.  This includes a discussion of important features of slowly evolving orbits.  We then discuss adiabatic inspiral in Sec.\ \ref{sec:adiabinsp}, with the details of inspiral laid out in Sec.\ \ref{sec:inspiral}, and our criteria for ending inspiral discussed in Sec.\ \ref{sec:endofinspiral}.  Our diagnostic for the end of inspiral relies on comparing two notions of acceleration: the coordinate acceleration at periapsis associated with geodesic motion (using the Mino-time parameterization), and the rate at which the periapsis location accelerates (again using Mino time) due to GW backreaction.

We turn to a discussion of the transition from inspiral to plunge in Sec.\ \ref{sec:trans}.  As described above, with eccentricity there is no universal solution describing this epoch of the system's evolution, in strong contrast to what has been found for quasi-circular inspiral.  Instead, two systems which are indistinguishable except for their initial relativistic anomaly can have quite different transition and plunge behaviors.  In this way, the late evolutionary behavior of eccentric systems significantly differs from the late behavior of circular systems.  In Sec.\ \ref{sec:worldline}, we discuss how to develop the full coordinate-space worldline from our inspiral-transition-plunge solution, and we investigate the results we find in Sec.\ \ref{sec:results}.  We begin our discussion (Sec.\ \ref{subsec:ResultsA}) by looking at a representative example of a worldline constructed with our prescription.  We then study how the choice of initial radial phase affects worldlines in Sec.\  \ref{subsec:ResultsB}, demonstrating the key role that this angle plays on the binary's late-time dynamics.

In Sec.\ \ref{subsec:ResultsB}, we also examine our model's sensitivity to two {\it ad hoc} parameters we introduce: one, $\alpha_{\rm IT}$, which determines when the acceleration due to GW backreaction is ``strong enough'' to end inspiral and begin the transition; and another, $\beta_{\rm TP}$, which controls when we change our transition to a plunging geodesic.  These parameters are defined explicitly in Sections \ref{sec:endofinspiral} and \ref{sec:trans}, respectively.  We find a notably weak dependence on $\beta_{\rm TP}$, indicating that our models appear to be robust against the way we choose to end transition and begin the plunge.  By contrast, we find a strong dependence on $\alpha_{\rm IT}$ unless we confine it to a subset of its range of plausible values. It should be emphasized that we do not have strong physical reasons for choosing particular values for either of these parameters.  We expect that work carefully examining how eccentric inspiral ceases to evolve adiabatically, akin to the analyses of Refs.\ \cite{CompereKuchler2021, kuchler2024} for quasi-circular inspiral, will play an important role in understanding how to better model these sources through this regime.

In our conclusions, Sec.\ \ref{sec:conclude}, we outline plans to extend this framework to generic (inclined and eccentric) orbital geometries, and to use this transition framework to investigate GWs produced from systems with substantial eccentricity at the end of inspiral.  A major goal of our ongoing study is to examine how the final GWs emitted by a small-mass-ratio binary system depend upon the system's eccentricity as inspiral comes to an end.  A diagnostic of eccentricity in the late waveform would be a valuable tool for understanding properties of binaries which produce these signals, especially for massive systems for which only the final few wave cycles are in GW detectors' sensitive bands.

We use geometrized units with $G = 1$, $c = 1$; a useful conversion factor in these units is $1\,M_\odot = 4.93\times 10^{-6}\,{\rm sec}$.  The larger black hole is described by the Kerr metric, with mass $M$ and spin parameter $a$.  The smaller body has mass $\mu$, and the mass ratio is $\eta = \mu/M$.

\section{Kerr geodesic orbits}
\label{sec:geods}

We treat inspiral as a slowly-evolving sequence of geodesic orbits, and the transition from inspiral to plunge is based on a modification of geodesic motion.  We thus begin by briefly reviewing keye characteristics of Kerr black hole geodesic orbits.  This material has been presented in depth elsewhere \cite{MTW, Schmidt2002, DrascoHughes2004, FujitaHikida, vandeMeent2020}; our review is largely to introduce important concepts and quantities for our analysis, as well as to introduce notation. After reviewing Kerr geodesics, we consider the conditions under which ``slowly evolving'' orbits exist.  This discussion is similar in setup to previous discussion of slowly evolving quasi-circular orbits --- see Sec.\ IIIC of Ref.\ \cite{ApteHughes2019}; a brief summary is given in Appendix A of this paper.  The focus on eccentric orbits significantly changes one aspect of this analysis.

\subsection{Generalities}

In Boyer-Lindquist coordinates $(t,r,\theta,\phi)$, a geodesic orbit in Kerr spacetime is specified by the equations 
\begin{eqnarray}
\left(\frac{dr}{d\lambda}\right)^2 &=& \left[E(r^2+a^2) - a L_z\right]^2
\nonumber\\
& & - \Delta\left[r^2 + (L_z - a E)^2 + Q\right]
\nonumber\\
&\equiv& R(r)\;,\label{eq:rdot}\\
\left(\frac{d\theta}{d\lambda}\right)^2 &=& Q - \cot^2\theta L_z^2 -a^2\cos^2\theta(1 - E^2)
\nonumber\\
&\equiv& \Theta(\theta)\;,\label{eq:thetadot}\\
 \frac{d\phi}{d\lambda} &=& \csc^2\theta L_z + \frac{2Mra E}{\Delta} -\frac{a^2 L_z}{\Delta}
\nonumber\\
&\equiv& \Phi(r,\theta)\;,\label{eq:phidot}\\
\frac{dt}{d\lambda} &=& E\left[\frac{(r^2 + a^2)^2}{\Delta} - a^2\sin^2\theta\right] - \frac{2Mra L_z}{\Delta}
\nonumber\\
&\equiv& T(r,\theta),\;\label{eq:tdot}
\end{eqnarray}
where 
\begin{equation}
    \Delta \equiv r^2 - 2Mr + a^2\;.
\end{equation}
See, for example, Eqs.\ (33.32a--d) of \cite{MTW}.  The quantity $E$ is the orbital energy (per unit $\mu$), $L_z$ is the axial angular momentum (per unit $\mu$) and $Q$ is the orbit's Carter constant (per unit $\mu^2$).  These are integrals of motion that specify a family of geodesic orbits, up to initial conditions.  Along a geodesic, $E$, $L_z$ and $Q$ are conserved.  Kerr geodesics also depend on the black hole spin parameter $a$, which gives the angular momentum per unit mass of the primary and lies in the range $0 \le a \le M$.  We distinguish between prograde and retrograde orbits with the sign of $L_z$.  The parameter $\lambda$ is Mino time, a time parameter that is particularly well-suited for describing orbits in Kerr spacetime.  An interval of $\lambda$ is related to an interval of proper time along the orbit by $d\tau = \Sigma\,d\lambda$, where $\Sigma \equiv r^2 + a^2\cos^2\theta$.

The Boyer-Lindquist coordinate $t$ describes time as measured by distant observers.  We use Eq.\ (\ref{eq:tdot}) to convert between intervals of Mino time and intervals of observer time.  When discussing adiabatic evolution, it is often useful to orbit average the function $T(r,\theta)$:
\begin{equation}
    \Gamma = \left\langle T(r,\theta)\right\rangle\;.
    \label{eq:Gamma}
\end{equation}
See Ref.\ \cite{DrascoHughes2004} for a precise definition of this averaging, and Ref.\ \cite{FujitaHikida} for analytic expressions of the resulting integrals.  The factor $\Gamma$ converts orbit-averaged quantities expressed in a per Mino-time basis to orbit-averaged quantities expressed in a per coordinate-time basis.

Several important properties of an orbit can be deduced by examining $R(r)$:
\begin{eqnarray}
R(r) &=& (E^2 - 1)r^4 + 2M r^3 +[a^2(E^2 - 1) - L_z^2 - Q]r^2
\nonumber\\
& & + 2M[Q + (aE - L_z)^2]r - a^2Q\;,
\label{eq:quartic}\\
&=& (1 - E^2)(r_{\rm a} - r)(r - r_{\rm p})(r - r_3)(r - r_4)\;.
\label{eq:Rroots}
\end{eqnarray}
The form (\ref{eq:quartic}) is found by expanding Eq.\ (\ref{eq:rdot}); the form (\ref{eq:Rroots}) reorganizes the resulting quartic, expressing $R(r)$ as a function of its four roots.  When $\{E, L_z, Q\}$ corresponds to an eccentric, bound Kerr orbit, the geodesic oscillates over the radial domain $r_{\rm p} \le r \le r_{\rm a}$.  The Mino-time period $\Lambda_r$ describing this oscillation is given by 
\begin{equation}
\Lambda_r = 2\int_{r_{\rm p}}^{r_{\rm a}}\frac{dr}{\sqrt{R(r)}}\;.
   \label{eq:RadialPeriod} 
\end{equation}
This is converted to a Boyer-Lindquist-time period using the factor $\Gamma$ defined in Eq.\ (\ref{eq:Gamma}):
\begin{equation}
    T_r = \Gamma\,\Lambda_r\;;
\end{equation}
$T_r$ is the period of radial motion that is measured by distant observers.  A similar period $T_\phi$ is associated with motion in the axial direction, and for general orbits a period $T_\theta$ describes oscillations in the polar direction.  Further discussion and detailed formulas can be found in Refs.\ \cite{DrascoHughes2004, FujitaHikida}.  Our analysis focuses on $T_r$.

The roots are ordered such that $r_{\rm a} \ge r_{\rm p} \ge r_3 > r_4$ for bound orbits.  The root $r_4$ is generally inside the event horizon, and takes the value $r_4 = 0$ for Schwarzschild black holes ($a = 0$) and for equatorial orbits ($Q = 0$).  It is common to remap the roots $r_{\rm p}$ and $r_{\rm a}$ to parameters $p$ and $e$ according to
\begin{equation}
    r_{\rm p} = \frac{p}{1 + e}\;,\quad r_{\rm a} = \frac{p}{1 - e}\;.
\end{equation}
A useful parameterization of the radial motion builds in these turning points by writing
\begin{equation}
    r = \frac{p}{1 + e\cos(\chi_r + \chi_{r0})}\;.
    \label{eq:chir}
\end{equation}
As the relativistic anomaly angle $\chi_r$ increases from $0$ to $2\pi$, the radial motion oscillates through a full cycle, covering periapsis $r_{\rm p}$ to apopasis $r_{\rm a}$ and back.  We typically align $\chi_r$ such that $\chi_r = 0$ when $\lambda = 0$; the parameter $\chi_{r0}$ then sets an orbit's initial radial position.  By substituting Eq.\ (\ref{eq:chir}) into Eq.\ (\ref{eq:rdot}), we find an equation for $d\chi_r/d\lambda$ that is particularly useful for our applications.  See Appendix A of Ref.\ \cite{DrascoHughes2004} for detailed discussion.

For general Kerr geodesics, the orbit's inclination angle $I$ is related to its Carter constant $Q$ (as well as to the energy $E$ and axial angular momentum $L_z$), leading to motion in which the polar angle $\theta$ oscillates about the equatorial plane.  We restrict our detailed analysis in what follows to equatorial orbits, for which $\theta = \pi/2$ at all times, and thus $I = 0^\circ$ or $180^\circ$.  We discuss in our conclusions plans to generalize our approach to include inclined orbits, and refer the reader to AH2019 for detailed discussion of a transition with inclination in the circular limit. Equatorial orbits have $Q = 0$, so we also set the Carter constant to zero in this work.
 
These considerations tell us that the parameter sets $\{E, L_z, Q, a\}$ and $\{a, p, e, I\}$ are equivalent and interchangeable; indeed, formulas given in Refs.\ \cite{FujitaHikida, vandeMeent2020, Hughes2024} implement mappings between these different forms.  In this paper, we treat these ways of parameterizing an orbit as identical, switching between them depending on which is most useful for the analysis at hand.

\subsection{Slowly evolving orbits}
\label{sec:slowlyevolving}

When examining how orbits evolve due to GW backreaction, it is quite useful to consider the competing timescales governing the dynamics: the ``short" orbital timescale associated with the period of the eccentric orbit, $T_r$, and the ``long" timescale on which the orbit decays due to radiation reaction, $T_{\rm RR}$.  Independent of the chosen time parameter ($i.e.$, whether we use Mino time, coordinate time, or proper time), the ratio of these timescales is given by the mass ratio
\[
\frac{T_r}{T_{\rm RR}}\sim \eta\;.
\]
In the slowly-evolving limit, in which the inspiral is well-described by the small body progressing through a sequence of geodesics, this separation of timescales allows us to rewrite our equation of motion and further constrain the dynamics.  Recall the Kerr radial equation of motion, Eq.\ (\ref{eq:rdot}):
\begin{equation}
 \fpar{dr}{d\lambda}^2 = R(r)\;.
\label{eq:rdot2}
\end{equation}
The function $R(r)$ also depends on the integrals of motion $E, L_z$ and $Q$. These quantities are constant on a geodesic, but slowly evolve when we include backreaction.  Apply $d/d\lambda$ to both sides of Eq.\ (\ref{eq:rdot2}):
\begin{eqnarray}
2\fpar{dr}{d\lambda}\fpar{d^2r}{d\lambda^2} &=& \f{\partial R}{\partial E}\fpar{dE}{d\lambda}
+  \f{\partial R}{\partial L_z}\fpar{dL_z}{d\lambda}
\nonumber\\
&+& \f{\partial R}{\partial Q}\fpar{dQ}{d\lambda} + \frac{\partial R}{\partial r}\fpar{dr}{d\lambda}\;.
\nonumber\\
\label{eq:Rderiv}
\end{eqnarray}
Now examine how these various terms scale with the mass ratio $\eta$:
\begin{itemize}

\item The GW-driven rates of change $dE/d\lambda$, $dL_z/d\lambda$ and $dQ/d\lambda$ are all of order $\eta$ (bearing in mind that $E$ and $L_z$ are themselves expressed per unit mass $\mu$, and $Q$ is per unit $\mu^2$).

\item The derivatives $\partial R/\partial E$, $\partial R/\partial L_z$ and $\partial R/\partial Q$  are all independent of $\eta$.

\item The derivative $\partial R/\partial r$ is independent of $\eta$ and tends to be large, at least away from the LSO.

\item For eccentric orbits, both the Mino-time radial velocity $dr/d\lambda$ and the acceleration $d^2r/d\lambda^2$ are dominantly independent of mass ratio, though both acquire corrections of $\mathcal{O}(\eta)$ due to backreaction.

\end{itemize}
The final two items in this list are quite different from the equivalent statements describing radial motion for quasi-circular inspiral and transition (see Appendix A for details of the quasi-circular case).  In the quasi-circular limit, $\partial R/\partial r$ is zero for geodesics, as are the radial velocity $dr/d\lambda$ and radial acceleration $d^2r/d\lambda^2$.  Each of these quantities acquires $\mathcal{O}(\eta)$ corrections due to backreaction, which dominates their behavior in the quasi-circular case.  By contrast, for eccentric motion, each of these terms is dominated by their non-zero, geodesic values, which are independent of the mass ratio.

Relating these scaling relationships to Eq.\  (\ref{eq:Rderiv}) reveals that some terms are $\mathcal{O(\eta)}$, while others are $\mathcal{O}(1)$.  Equation (\ref{eq:Rderiv}) must hold at each order of $\eta$, which allows us to separate Eq.\ (\ref{eq:Rderiv}) as follows:
\begin{equation}
\mathcal{O}(\eta):\;\;\;\f{\partial R}{\partial E}\fpar{dE}{d\lambda}+\f{\partial R}{\partial L_z}\fpar{dL_z}{d\lambda} + \f{\partial R}{\partial Q}\fpar{dQ}{d\lambda} = 0\;,  
\label{eq:etaOrderconstraint}    
\end{equation}
\begin{equation}
\mathcal{O}(1):\;\;\; \frac{d^2r}{d\lambda^2} = \f{1}{2} \frac{\partial R}{\partial r}\;.
\label{eq:acceleqn} 
\end{equation}
The $\mathcal{O}(\eta)$ relation, Eq.\ (\ref{eq:etaOrderconstraint}), constrains the evolution of the integrals of motion in the slowly-evolving limit.  When $E, L_z$ and $Q$ are constant, Eq.\ (\ref{eq:etaOrderconstraint}) is trivially satisfied, and Eq.\ (\ref{eq:acceleqn}) is simply a rewriting of the radial part of the geodesic equation in Kerr spacetime.  As the LSO is approached, we expect $\mathcal{O}(\eta)$ corrections to geodesic behavior arising from the evolution of the integrals of motion to become increasingly important.

\section{Adiabatic inspiral with eccentricity, and how it ends}
\label{sec:adiabinsp}

We begin this discussion by reviewing how we compute adiabatic inspirals, and examining when adiabatic inspiral has come to an end.  We diagnose the adiabaticity of inspiral by comparing two notions of acceleration: one associated with an orbit's radial coordinate motion, and another arising due to the backreaction of GW emission.

\subsection{Adiabatic inspiral}
\label{sec:inspiral}

Here we describe how we compute adiabatic inspiral in more detail, examining in particular the conditions under which adiabatic evolution holds.  We do this for general spin parameter $a$, though we focus on equatorial orbits ($Q = 0$, $\cos I = \pm 1$).  Generalization to inclined orbits should be straightforward.  We discuss how we expect this generalization to proceed in our conclusions, Sec.\ \ref{sec:conclude}.

We begin our computation of adiabatic inspiral by developing a set of backreaction data that covers our parameter space.  For each spin that we study, our data is on a grid which breaks into two subgrids: an ``outer'' grid that covers a region somewhat outside the LSO to moderately large separation, and an ``inner'' grid that densely samples a region close to the LSO.  The outer grids we use were originally developed for studies of a Kerr extension of the Fast EMRI Waveform (``FEW'') project \cite{Chua2021, Katz2021, Speri2023}.  These grids cover the domain $0 \le e \le 0.8$, with 36 points uniformly spaced in $e^2$ in order to more densely cover small eccentricity, as well as 40 points in $p$ spaced according to the formula
\begin{equation}
    p^{\rm out}_j = p^{\rm out}_{\rm min}(a, e) + 4M\left(e^{j\Delta u} - 1\right)\;,\quad 0 \le j \le 39\;.
    \label{eq:outgrid}
\end{equation}
For these data sets\footnote{It should be emphasized that the grids which we use here are not the final versions that are being used for the Kerr extension of FEW, which will be presented elsewhere.}, we use $p^{\rm out}_{\rm min}(a, e) = p_{\rm LSO}(a, e) + 0.05M$, and $\Delta u = 0.035$.

\begin{figure}[ht]
\includegraphics[scale=0.415]{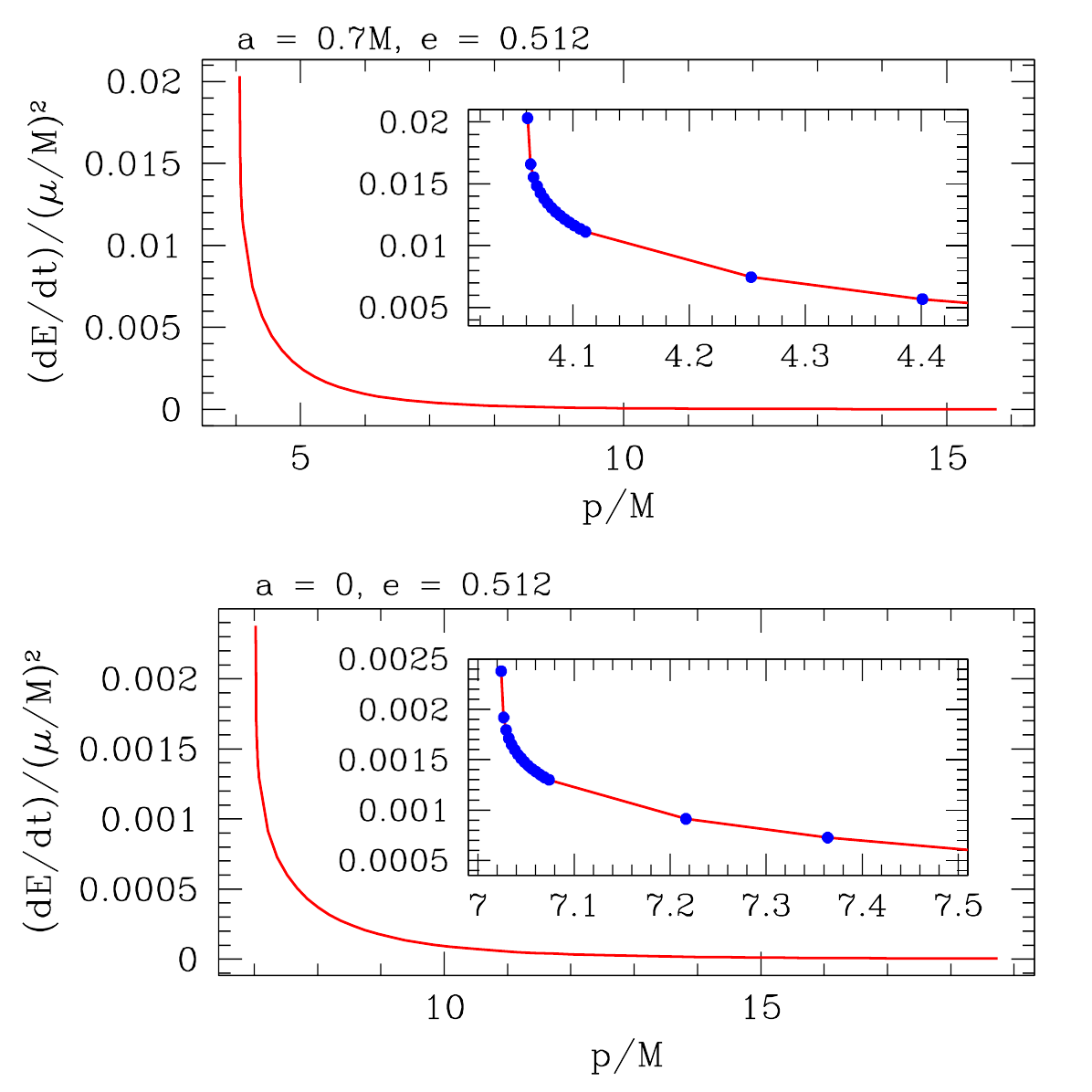}
\caption{Example of how fluxes behave on our grid near the LSO.  Top panel shows $dE/dt$ for $a = 0.7M$, $e = 0.512$, for which $p_{\rm LSO} = 4.061M$.  Bottom panel shows $dE/dt$ for $a = 0$, $e = 0.512$, for which $p_{\rm LSO} = 7.024M$.  Notice in both cases that the flux rises quickly as we approach the LSO.  The inset panels zoom in on the region closest to the LSO, showing (blue dots) the higher-density data we compute to ensure we have accurate interpolations in this region.  We place the inner edge of our grid at $p_{\rm LSO} + 10^{-4}M$, though we have verified that all flux data are well behaved all the way to $p = p_{\rm LSO}$.}
\label{fig:e0.512_flux}
\end{figure}

Our model for the transition from inspiral to plunge requires adiabatic backreaction data from a region very close to the LSO.  Quantities which describe backreaction vary rather sharply over the domain $p_{\rm LSO}(a, e) \le p \le p_{\rm LSO}(a, e) + 0.05M$.  In order to construct accurate interpolants over this region, we develop a small but dense inner subgrid covering the region near the LSO:
\begin{equation}
    p^{\rm in}_j = p^{\rm in}_{\rm min}(a, e) + \Delta p(e)\left(e^{j(\ln 2)/14} - 1\right)\;,\quad 0\le j \le 14\;.
    \label{eq:ingrid}
\end{equation}
We use $p^{\rm in}_{\rm min}(a, e) = p_{\rm LSO}(a, e) + 10^{-4}M$, and $\Delta p(e) = p^{\rm out}_{\rm min}(a, e)- p^{\rm in}_{\rm min}(a, e)$.  At each orbit, we compute $(dE/dt, dL_z/dt)$ using the Teukolsky equation \cite{Teukolsky1973} in a frequency-domain decomposition, including enough modes that fluxes converge\footnote{Convergence at fractional accuracy $10^{-5}$ is adequate for our studies, though not for other work.  For example, the data sets used for FEW require more accuracy.} to a part in $10^{-5}$.  See \cite{Hughes2021} for a detailed discussion of the algorithm used to compute fluxes.  Figure \ref{fig:e0.512_flux} shows examples of the fluxes we use on our grid, and in particular how flux data rapidly vary as the LSO is approached.  This steep growth appears to reflect the ``zoom-whirl'' character of orbits as they approach the LSO; the peak in the fluxes behaves in a simple way as a function of the number of axial revolutions the orbit makes for each full radial cycle.  This behavior will be discussed in detail elsewhere.

\begin{figure*}[ht]
\includegraphics[scale=0.415]{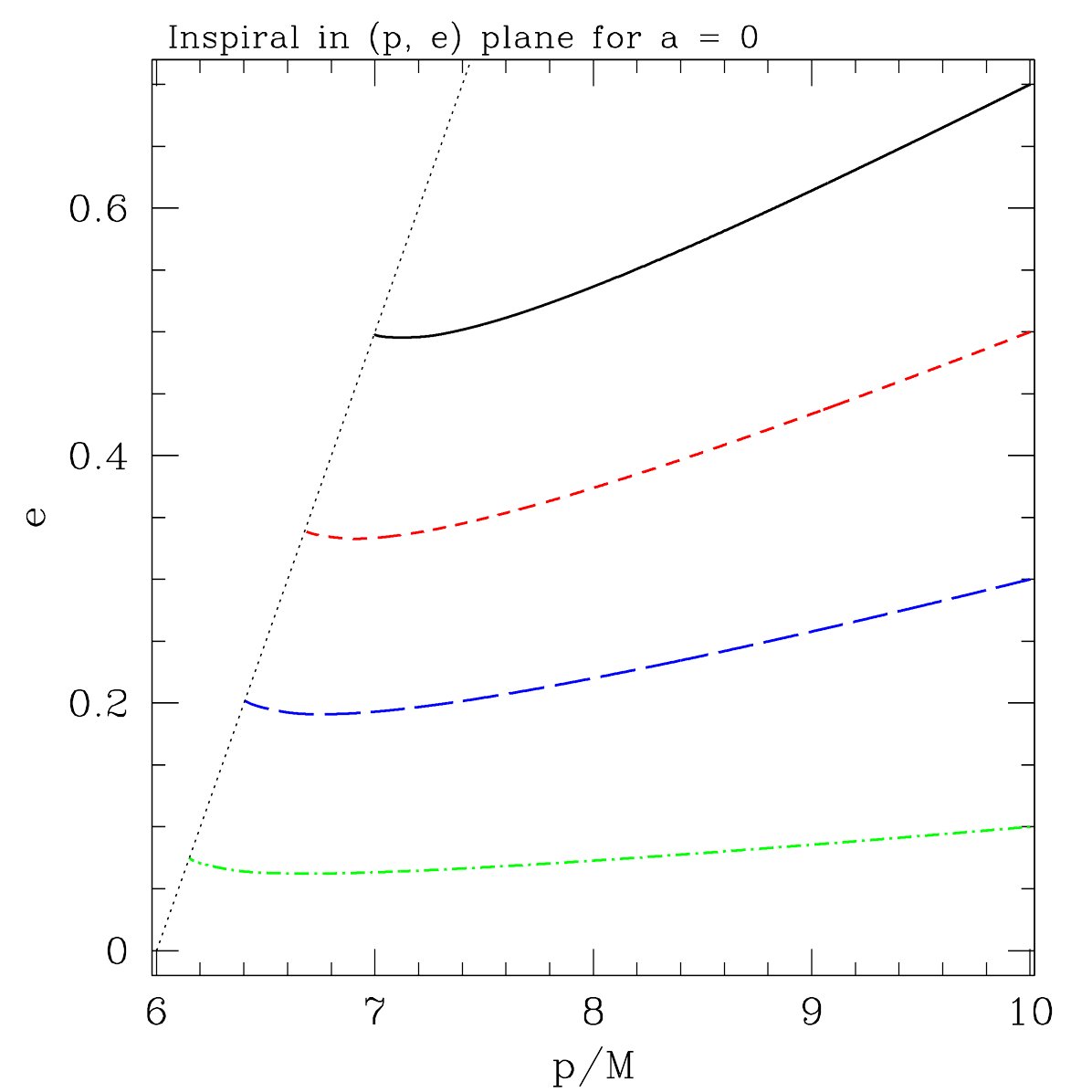}
\hfill
\includegraphics[scale=0.415]{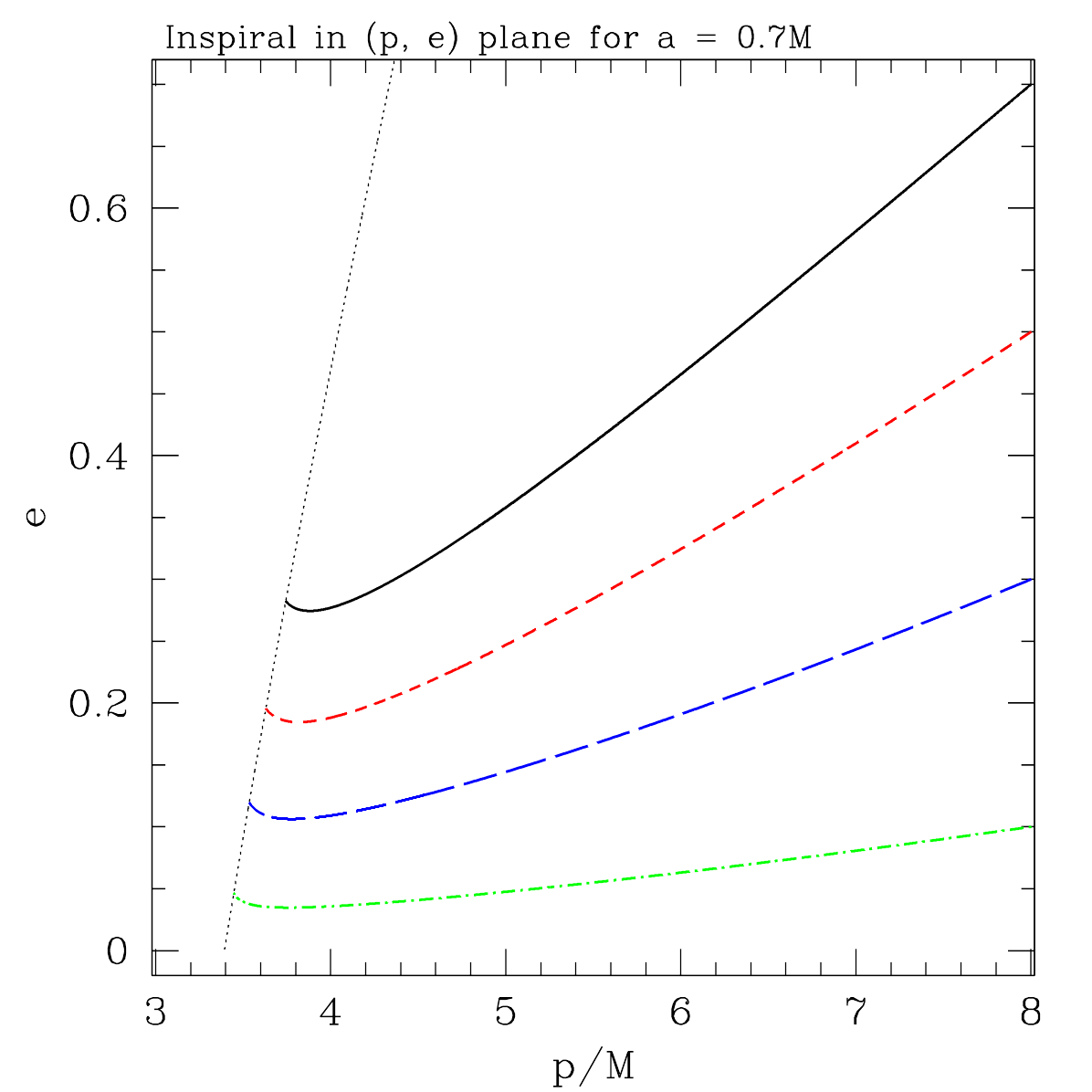}
\caption{Adiabatic inspiral into a Schwarzschild black hole (left panel) and into a Kerr black hole with spin parameter $a = 0.7M$ (right).  Both panels show the trajectory followed by adiabatic inspiral in the $(p, e)$ plane, starting at $p = 10M$ for Schwarzschild, and $p = 8M$ for Kerr.  In all cases, inspiral is taken to the LSO (dotted line).  Four examples are shown for each black hole spin, starting at $e = 0.7$ (solid black curve), $e = 0.5$ (short-dashed red curve), $e = 0.3$ (long-dashed blue curve), and $e = 0.1$ (dashed-dotted green curve).  In the adiabatic limit, these curves are independent of the binary's masses, though each inspiral's duration scales inversely with mass ratio (e.g., the inspirals starting at $e = 0.7$ last $T_{\rm insp} = 94.42 M/\eta$ for $a = 0$, and $T_{\rm insp} = 90.20 M/\eta$ for $a = 0.7M$).}
\label{fig:a0.0_a0.7_insp}
\end{figure*}

To construct an inspiral, we start at an initial point $(p_i, e_i)$ in this plane.  From this point, we integrate up $(dE/dt, dL_z/dt)$ to construct $[E(t), L_z(t)]$ along the inspiral.  Using Ref.\ \cite{Hughes2024}, it is simple to use this to build the trajectory $[p(t), e(t)]$ that the adiabatic inspiral follows from $(p_i, e_i)$ to a point $(p_f, e_f)$ just outside the LSO.  The duration of such an inspiral scales as $M/\eta = M^2/\mu$.  Figure \ref{fig:a0.0_a0.7_insp} shows examples of inspiral in the $(p, e)$ plane for $a = 0$ and $a = 0.7M$.

\subsection{Diagnosing the end of adiabatic inspiral}
\label{sec:endofinspiral}

As the secondary enters the strong field, gravitational backreaction increasingly dominates the dynamics, eventually invalidating our assumption of adiabatic evolution.  Here we describe how we assess when adiabatic inspiral has ended by comparing two notions of acceleration that describe an eccentric inspiral.

We begin with the geodesic limit.  A body moving on an eccentric geodesic experiences a coordinate acceleration that oscillates over its orbit, outward directed at periapsis and inward directed at apoapsis.  As shown in Sec.\ \ref{sec:slowlyevolving}, the coordinate radial acceleration of a body moving on a geodesic is given by
\begin{equation}
 \frac{d^2r}{d\lambda^2} = \frac{1}{2}\frac{\partial R}{\partial r}\;.
 \label{eq:accel2}
\end{equation}
Evaluating this at periapsis $r_{\rm{p}} = p/(1+e)$, we find
\begin{eqnarray}
    a_{\rm p, geod} \equiv \left.\frac{d^2r}{d\lambda^2}\right|_{r = r_{\rm p}} &=& \left(L_z - aE\right)^2 M
    \nonumber\\
    &-& \frac{p\left(L_z^2 + a^2(1 - E^2)\right)}{(1 + e)}
    \nonumber\\
    &+& \frac{3Mp^2}{(1 + e)^2} - \frac{2(1 - E^2)p^3}{(1 + e)^3}\;.
    \nonumber\\
    \label{eq:periaccel_out}
\end{eqnarray}
The Schwarzschild limit of (\ref{eq:periaccel_out}) is particularly clean, since both $E$ and $L_z$ have simple forms when $a = 0$:
\begin{equation}
    a_{\rm p, geod}\bigr|_{a \to 0} = \frac{p^2eM\left(p - (6+2e)M\right)}{(1 + e)^2\left(p - (3+e^2)M\right)}\;.
\end{equation}
Notice that this vanishes at the Schwarzschild LSO, $p = (6+2e)M$.  Though it is not obvious in the general form (\ref{eq:periaccel_out}), the vanishing of $a_{\rm p, geod}$ at the LSO holds for all black hole spins.

Now consider how GW emission backreacts on the orbit, in particular how this backreaction affects the inner turning point $r_{\rm p}$.  Thanks to this backreaction, the inner turning point acquires a coordinate velocity
\begin{equation}
    v_{\rm p,GW} \equiv \frac{dr_{\rm p,GW}}{d\lambda} = \frac{J_{1E}(r_{\rm p})}{\mathcal{D}(r_{\rm p})}\frac{dE}{d\lambda} + \frac{J_{1L_z}(r_{\rm p})}{\mathcal{D}(r_{\rm p})}\frac{dL_z}{d\lambda}\;.
\label{eq:rpInwardvelocity}
\end{equation}
(We have omitted a term involving $dQ/d\lambda$ which is zero for equatorial orbits.)  The rates of change $dE/d\lambda$ and $dL_z/d\lambda$ that enter Eq.\ (\ref{eq:rpInwardvelocity}) are determined from our grid of adiabatic radiation reaction data, as discussed in Sec.\ \ref{sec:adiabinsp}.  We convert from rates of change per unit Boyer-Lindquist coordinate time to rates of change per unit Mino time in an orbit-averaged sense using
\begin{equation}
    \frac{dE}{d\lambda} = \Gamma \frac{dE}{dt}\;,\quad
    \frac{dL_z}{d\lambda} = \Gamma \frac{dL_z}{dt}\;,
\end{equation}
where the conversion factor $\Gamma$ is given by Eq.\ (\ref{eq:Gamma}).  The quantities $J_{1E}(r_{\rm p})$, $J_{1L_z}(r_{\rm p})$ and $\mathcal{D}(r_{\rm p})$ appearing in Eq.\ (\ref{eq:rpInwardvelocity}) are taken from the Jacobian which relates the rate of change of orbital geometry to the rate of change of integrals of the motion, evaluated at $r = r_{\rm p}$.  Expressions for these quantities can be read out of Eqs.\ (B5) and (B6) of \cite{Hughes2021}, specializing to $Q = 0$:
\begin{eqnarray}
    J_{1E}(r) &=& 4aM(L_z - aE) r - 2Er^2(r^2 + a^2)\;,
    \label{eq:J1Erp}\\
    J_{1L_z}(r) &=& 4M(aE - L_z)r + 2L_z r^2\;,
    \label{eq:J1Lzrp}\\
    \mathcal{D}(r) &=& 2M(L_z - aE)^2 - 2r\left(L_z^2 + a^2(1 - E^2)\right)
    \nonumber\\ & &
    + 6Mr^2 - 4(1 - E^2)r^3\;.
    \label{eq:JacobianDenom}
\end{eqnarray}
Since $E$ and $L_z$ are computed per unit orbiting body mass, the Jacobian elements are all independent of the secondary's mass.  Additionally, $dE/d\lambda$ and $dL_z/d\lambda$ both scale with the reduced mass ratio $\eta$.  The inward speed $v_{\rm p, GW}$ is thus proportional to $\eta$.

Applying an additional Mino-time derivative to Eq.\ (\ref{eq:rpInwardvelocity}) tells us the acceleration of $r_{\rm p}$ due to GW-driven backreaction:
\begin{widetext}
\begin{eqnarray}
    a_{\rm p, GW} \equiv \frac{d^2r_{\rm p, GW}}{d\lambda^2} &=& \frac{dr_{\rm p}}{d\lambda}\left[\frac{\partial}{\partial r}\left(\frac{dr_{\rm p}}{d\lambda}\right)\right]\biggl|_{r = r_{\rm p}}
    \nonumber\\
    &=& v_{\rm p, GW}\biggl[\biggl(\frac{J'_{1E}(r_{\rm p})}{\mathcal{D}(r_{\rm p})} - \frac{J_{1E}(r_{\rm p})\mathcal{D}'(r_{\rm p})}{\mathcal{D}(r_{\rm p})^2}\biggr) \frac{dE}{d\lambda} +
    \biggl(\frac{J'_{1L_z}(r_{\rm p})}{\mathcal{D}(r_{\rm p})} - \frac{J_{1L_z}(r_{\rm p})\mathcal{D}'(r_{\rm p})}{\mathcal{D}(r_{\rm p})^2}\biggr) \frac{dL_z}{d\lambda}\biggr]\;.
    \label{eq:periaccel_in}
\end{eqnarray}
\end{widetext}
Prime denotes $\partial/\partial r$, so (for example) $J'_{1E}(r_{\rm p})$ means $\partial J_{1E}/\partial r$ evaluated at $r = r_{\rm p}$.

Because $v_{\rm p,GW}$, $dE/d\lambda$, and $dL_z/d\lambda$ all scale with $\eta$, while the Jacobian elements are each independent of mass ratio, $a_{\rm p, GW}$ scales with $\eta^2$.  This scaling indicates that $a_{\rm p, GW}$ is quite small, at least until orbits are deep within the strong field, where radiation reaction is most intense.  In the strong-field region where $a_{\rm p, GW}$ is largest, $a_{\rm p, geod}$ becomes small since we are near the LSO.  As $a_{\rm p, GW}$ approaches $a_{\rm p, geod}$, backreaction becomes so strong that the system is no longer in the regime of adiabatic inspiral, but has begun the transition to plunge.

\begin{figure*}[ht]
\includegraphics[scale=0.415]{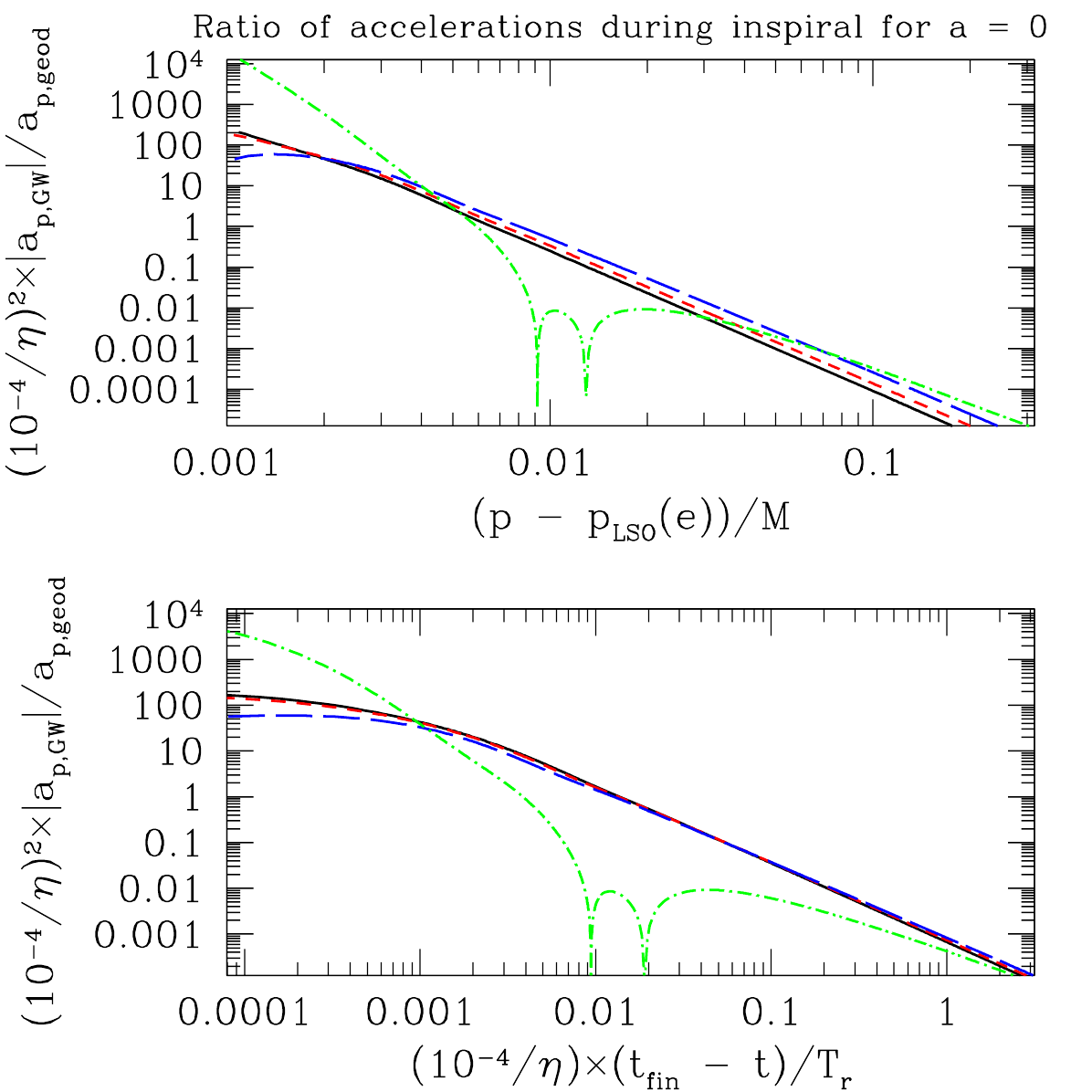}
\hfill
\includegraphics[scale=0.415]{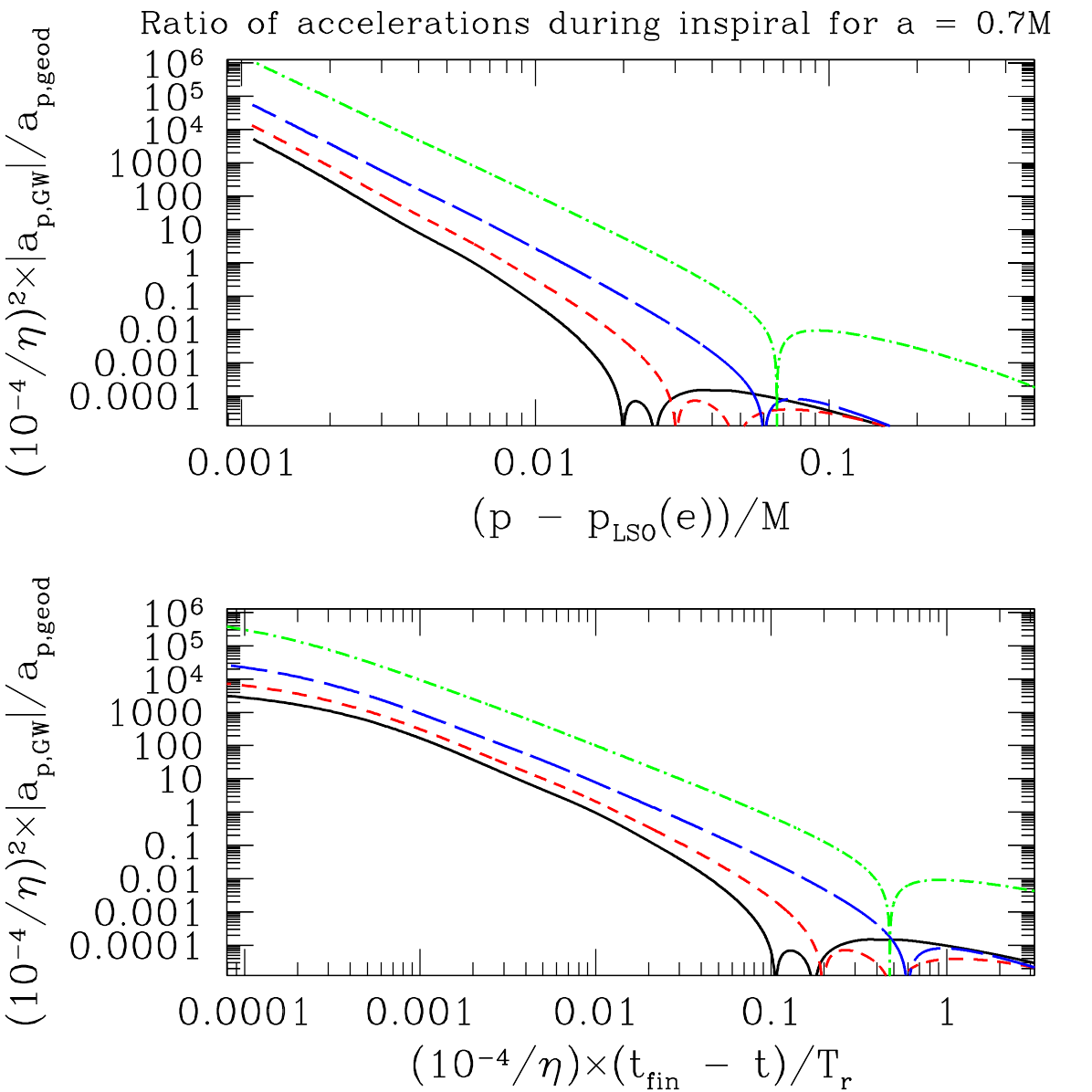}
\caption{Comparison of acceleration of the periapsis due to GW backreaction, $a_{\rm p,GW}$, to the secondary body's geodesic coordinate acceleration at this radius, $a_{\rm p,geod}$.  Left-hand panels are for inspiral into a Schwarzschild black hole; right shows inspiral into a Kerr black hole with $a = 0.7M$.  Top panels plot this ratio as a function of separation from the LSO; bottom shows this as a function of time remaining until the end of adiabatic inspiral (normalized to the radial period).  In all panels, the inspirals shown are the same (with the same encoding of initial conditions) as those included in Fig.\ \ref{fig:a0.0_a0.7_insp}.  We find in all cases that $a_{\rm p,geod}$ is much larger in magnitude than $a_{\rm p,GW}$ until the inspiraling body is very close to the LSO, or equivalently within a fraction of a radial period of the end of adiabatic inspiral.  Oscillations in all Kerr cases and in the low eccentricity Schwarzschild case (the green curve in the left panels, where $e = 0.1$ at $p = 10M$) are due to $a_{\rm p,GW}$ switching sign in the strong field; we find similar oscillations across a range of eccentricities for prograde inspirals as the spin is increased from $0$ to $0.7M$.}
\label{fig:a0.0_a0.7_acc}
\end{figure*}

This behavior can be seen in Fig.\ \ref{fig:a0.0_a0.7_acc}, where we compare $a_{\rm p, GW}$ and $a_{\rm p, geod}$ for multiple adiabatic inspirals.  In all panels, we show $a_{\rm p, GW}/a_{\rm p, geod}$; the top panels examine this ratio versus separation from the LSO, the bottom panels do so versus time until the end of inspiral (normalized to the radial period $T_r$).  Panels on the left show data for $a = 0$; those on the right show $a = 0.7M$.  Each track corresponds to one of the inspirals shown in Fig.\ \ref{fig:a0.0_a0.7_insp}.  We see that $a_{\rm p, geod}$ is much larger in magnitude than $a_{\rm p, GW}$ until we are very close to the LSO, or equivalently a small fraction of a radial period from the end of inspiral.  The ratio of accelerations then rises rapidly, though in some cases it oscillates due to $a_{\rm p, GW}$ changing sign in the strong field.  For the cases shown in Fig.\ \ref{fig:a0.0_a0.7_acc}, we see oscillations for all Kerr inspirals; for Schwarzschild, we only see this for the smallest eccentricity examined.  We have examined inspirals for other black hole spins (not shown in order to reduce the number of figures), and find that oscillations in $a_{\rm p, GW}$ spread from small eccentricity to large as the black hole spin is increased.  These oscillations are most prominent for inspirals that go to relatively small values of $p$.

We define the onset of transition by comparing $a_{\rm p,geod}$ to $a_{\rm p,GW}$: inspiral ends and the transition begins when
\begin{equation}
    \f{a_{\rm p,GW}}{a_{\rm p, geod}} \ge \alpha_{\rm IT}\;.
\end{equation}
We have introduced a parameter $\alpha_{\rm IT}$ which is of order unity; when this ratio exceeds this parameter, we may be confident that backreaction is so strong that it is not accurate to treat the system as evolving through a sequence of geodesics.  In Sec.\ \ref{sec:results}, we explore how the worldlines we produce behave as $\alpha_{\rm IT}$ varies over the interval $[0.1, 1]$.

\section{Transition to plunge,\\ and how it ends}
\label{sec:trans}

Having described how we model the inspiral and how we decide when it has ended, we turn now to a discussion of the transition.  Our transition motion is found by directly integrating Eq.\ (\ref{eq:acceleqn}), but allowing the integrals of motion to evolve using a prescription which we describe below.  We also describe how we assess when transition has ended and the final plunge has begun.

\subsection{Transition solution}
\label{sec:transolution}

To describe the system during the transition, begin with the acceleration equation derived in Sec.\ \ref{sec:slowlyevolving}:
\begin{eqnarray}
\frac{d^2r}{d \lambda^2} &=& \frac{1}{2} \frac{\partial R}{\partial r}
\nonumber\\
&=& \left(L_z - aE\right)^2 M - \left(L_z^2 + a^2(1 - E^2)\right)r
\nonumber\\
& &+ 3Mr^2 - 2(1 - E^2)r^3\;.
\label{eq:radialgeo}
\end{eqnarray}
We have set $Q = 0$ in keeping with our focus on equatorial systems.  Following the framework originally developed in OT00 and refined in AH19, we continue to use Eq.\ (\ref{eq:radialgeo}) to describe the radial motion of the secondary, but now allow $E$ and $L_z$ to evolve due to gravitational radiation reaction.

For the quasi-circular analyses of OT00 and AH19, motion in the transition regime is described by expanding the function $R(r)$ in the radial variable $r$, as well as the integrals of motion $E$ and $L_z$ (and $Q$ in the case of AH19).  This expansion works well because the orbital radius is effectively an integral of the motion in the quasi-circular case, only changing by slowly shrinking due to GW emission.  In contrast, when the system is eccentric, the orbit's radius is dynamical, oscillating between $r_{\rm a}$ and $r_{\rm p}$ far more quickly than the timescale over which the integrals of motion evolve.

To describe an eccentric and equatorial transition, we therefore expand $R(r)$ only in $E$ and $L_z$.  Let
\begin{eqnarray}  
E &=&E^{\rm LSO} + \delta E\;,
\\
L_z &=& L_z^{\rm LSO} + \delta L_z\;.
\end{eqnarray}
The quantities $\delta E$ and $\delta L_z$ track the evolution of the integrals of motion through the transition.  We determine $\delta E$ and $\delta L_z$ using the prescription of AH2019, which takes as input information about adiabatic radiation reaction as described in Sec.\ \ref{sec:adiabinsp}.  Expanding $R(r)$ about the LSO, our equation of motion takes the form
\begin{equation}
    \frac{d^2r}{d\lambda^2} = \frac{1}{2}\left[\left(R'\right) + \left(\frac{\partial R'}{\partial E}\right)\delta E + \left(\frac{\partial R'}{\partial L_z}\right)\delta L_z\right]\;, 
    \label{eq:transitionPDE}
\end{equation}
where $R' = \partial R/\partial r$.  The right-hand side of this expression is a function of $r$, or of Mino-time $\lambda$; all quantities in parentheses are evaluated at the LSO.

We evolve the quantities $\delta E$ and $\delta L_z$ through the transition by using
\begin{equation}
    \mathcal{C} = \mathcal{C}_{\rm LSO}+\lambda\fpar{d\mathcal{C}}{d\lambda}_{\rm LSO}+\f{\lambda^2}{2}\mathcal{C}_2+\f{\lambda^3}{6}\mathcal{C}_3\;,
\label{eq:tranTimeEvolution}
\end{equation}
where $\mathcal{C} \in \{E, L_z\}$.  We define the origin $\lambda = 0$ to occur when the small body reaches the LSO\footnote{This definition requires us to offset time labels on the adiabatic inspiral trajectory in order to align with this choice of origin.}.  This corresponds to ``Model 2'' described in Sec.\ IV of AH2019; the corrections $\delta\mathcal{C}$ are the terms beyond $\mathcal{C}_{\rm LSO}$ in Eq.\ (\ref{eq:tranTimeEvolution}).  The quantities $\mathcal{C}_2$ and $\mathcal{C}_3$ are estimators for the second and third derivatives of $\mathcal{C}$ with respect to $\lambda$.  As shown in Eqs.\ (4.6) and (4.7) of AH2019, their values are constructed from $\mathcal{C}(\lambda_{\rm IT})$ and $(d\mathcal{C}/d\lambda)_{\lambda_{\rm IT}}$, where $\lambda_{\rm IT}$ denotes the Mino-time at which the transition begins\footnote{Note that AH2019 uses subscript $i$ to label the end of inspiral; $\lambda_i$ is thus the Mino time at which transition begins in that paper.}:
\begin{widetext}
\begin{align}
\mathcal{\mathcal{C}}_2 &= \f{2}{\lambda_{\rm IT}^2}\left[3\left(\mathcal{C}(\lambda_{\rm IT}) - \mathcal{C}_{\rm LSO}\right)- \lambda_{\rm IT}\left[2\fpar{d\mathcal{C}}{d\lambda}_{\rm LSO} + \fpar{d\mathcal{C}}{d\lambda}_{\lambda_{\rm IT}}\right]\right]\;,
\label{eq:C2}\\
\mathcal{\mathcal{C}}_3 &= \f{6}{\lambda_{\rm IT}^3}\left[2\left(\mathcal{C}_{\rm LSO} - \mathcal{C}(\lambda_{\rm IT})\right)+\lambda_{\rm IT}\left[\fpar{d\mathcal{C}}{d\lambda}_{\rm LSO}+\fpar{d\mathcal{C}}{d\lambda}_{\lambda_{\rm IT}}\right]\right].
\label{eq:C3}
\end{align}
\end{widetext}
The form (\ref{eq:tranTimeEvolution}), used with Eqs.\ (\ref{eq:C2}) and (\ref{eq:C3}), ensures that both the integrals of motion and their first time derivatives are continuous during the transition (correcting a discontinuity present in the original presentation of OT00).  This scheme for evolving the integrals of motion gives us all the information we need to integrate Eq.\ (\ref{eq:transitionPDE}) through the transition.  We do so using a fourth-order, fixed step size Runge-Kutta integrator, which is very simple and sufficiently accurate for the purpose of this study.

\subsection{End of transition and plunge}
\label{sec:endoftran}

We numerically integrate Eq.\ (\ref{eq:transitionPDE}) using the transition form (\ref{eq:tranTimeEvolution}) for the integrals of motion until the secondary reaches a coordinate distance $r_{\rm TP}$, defined by
\begin{equation}
r_{\rm TP} = r_{\rm H} + (1 - \beta_{\rm TP})(r_{\rm p, LSO} - r_{\rm H})\;.
\label{eq:blso}    
\end{equation}
Here $r_{\rm H} = M + \sqrt{M^2 - a^2}$ is the radius of the event horizon, $r_{\rm p, LSO}$ is the periapsis radius of the last stable orbit, and $\beta_{\rm TP}\in (0,1)$.  The parameter $\beta_{\rm TP}$ specifies the point between the LSO periapsis and the horizon where we switch from the transition to a plunging geodesic.  As $\beta_{\rm TP} \rightarrow 0$, the end of transition approaches the LSO; as $\beta_{\rm TP} \rightarrow 1$, the end of transition approaches the event horizon.  In Sec.\ \ref{subsec:ResultsB}, we explore how the worldlines we construct depend upon the choice of $\beta_{\rm TP}$ for several representative eccentricities.

Once we have reached $r_{\rm TP}$ and the transition has ended, we freeze the integrals of motion $E$ and $L_z$ and model the dynamics with a plunging geodesic.  We continue to describe the motion with Eq.\ (\ref{eq:acceleqn}), using $r_{\rm TP}$ and the value of $dr/d\lambda$ at that radius as starting conditions for the plunge.  We end the plunge when the secondary has crossed the event horizon radius $r_{\rm H}$.

Combining the inspiral, transition, and plunge gives us a complete solution for the radial motion of the binary's secondary from wide separation through the plunge into the larger black hole.  As we describe in the next section, we convert this into a full worldline by computing the motion in axial angle $\phi$ and Boyer-Lindquist time $t$ associated with this trajectory.

\section{Assembling the secondary's worldline in Mino and Boyer-Lindquist time}
\label{sec:worldline}

Applying the framework described in Sec.\ \ref{sec:trans}, we construct the radial trajectory that the secondary follows for adiabatic inspiral, transition, and plunge, parameterized by Mino time $\lambda$.  In this section, we use this $r(\lambda)$, along with the integrals of motion $E(\lambda)$ and $L_z(\lambda)$, to assemble the full coordinate-space worldline that this body follows, parameterized both in $\lambda$ and in Boyer-Lindquist coordinate time $t$.  Because the coordinate $t$ corresponds to time as measured by distant observers, this is a particularly useful variable when examining measurable features associated with the system.  For example, the worldline we assemble in this way is used to build the source term for the time-domain equation of black hole perturbation theory \cite{Teukolsky1973, Sundararajan2007, Sundararajan2008, Zenginoglu2011}; future work will use a code that solves this equation to study GWs produced by such systems.  We discuss our plans for such work in our conclusions.

Our tools for making the complete worldline and converting to time $t$ are the geodesic equations, Eqs.\ (\ref{eq:rdot})--(\ref{eq:tdot}), but replacing the constant integrals of motion with their evolving variants, $E(\lambda)$ and $L_z(\lambda)$.  We convert the time parameterization using
\begin{align}
    \frac{d\lambda}{dt} &= \frac{1}{T(r,\pi/2; \lambda)}
    \nonumber\\
    &= \frac{\Delta}{E(r^2 + a^2)^2 - 2aML_z r - \Delta a^2E}\;,
\end{align}
where $T(r,\theta; \lambda)$ is given in Eq.\ (\ref{eq:tdot}).  We use
\begin{align}
    \frac{d\phi}{dt} &\equiv \frac{d\phi}{d\lambda}\left(\frac{dt}{d\lambda}\right)^{-1}
    \nonumber\\
    &= \frac{2aME r - a^2L_z + \Delta L_z}{E(r^2 + a^2)^2 - 2aML_zr - \Delta a^2E}
\end{align}
to compute the axial angle $\phi$ along the worldline; in this expression, $r$, $E$, and $L_z$ are all taken to be functions of $\lambda$.  Notice that as $r \to r_{\rm H}$ (at which radius $\Delta = r^2 - 2Mr + a^2 \to 0$), $d\lambda/dt \to 0$, and
\begin{equation}
    \frac{d\phi}{dt} \to \frac{a}{2Mr_{\rm H}} \equiv \Omega_{\rm H}\;.
    \label{eq:OmegaH}
\end{equation}
The frequency $\Omega_{\rm H}$ is often called the angular velocity of the horizon.  An observer held just outside $r = r_{\rm H}$ orbits in $\phi$ at this rate as measured by distant observers.

These near-horizon features of $d\lambda/dt$ and $d\phi/dt$ enforce the well-known behavior that an infalling body ``freezes'' to the horizon from the perspective of a distant observer.  Although the worldline parameterized by $\lambda$ crosses the horizon, the worldline parameterized by $t$ does not.  Instead, as $t \to \infty$, $r(t) \to r_{\rm H}$, and the body ``whirls'' in the $\phi$ direction at the frequency $\Omega_{\rm H}$.

\section{Results}
\label{sec:results}

\subsection{Behavior of inspiral, transition, and plunge worldlines}
\label{subsec:ResultsA}

\begin{figure}[h!]
    \centering
    \includegraphics[scale = 0.4]{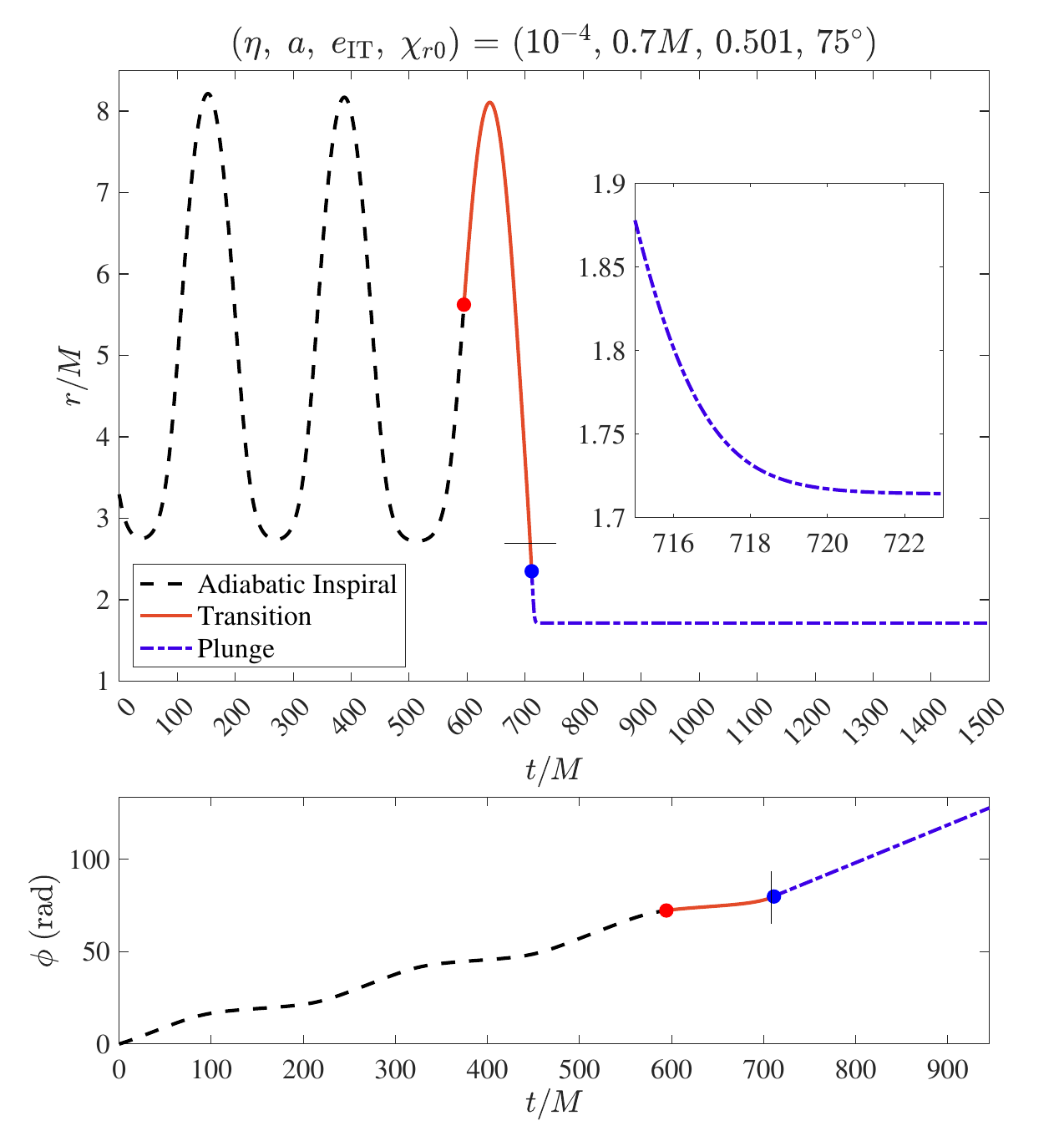}
    \caption{Example of a prograde inspiral-transition-plunge worldline constructed with our prescription for an eccentric binary with mass ratio $\eta = 10^{-4}$ and $a = 0.7M$. Top panel plots the orbital radius versus Boyer-Lindquist coordinate time; bottom panel shows the accumulation of the axial angle $\phi$ with time.  The span shown includes several cycles of adiabatic inspiral (the black, dashed line), ending at $t \approx 594M$, marked by the red dot.  (This time results from the parameter choice $\alpha_{\rm IT} = 0.80$, as defined in Sec.\ \ref{sec:endofinspiral}.)  The secondary then executes about three-quarters of a radial cycle during the transition, shown by the solid red curve.  It reaches the LSO at $t \approx 708M$ (designated by a black bar), continuing slightly beyond before we switch to a plunging geodesic at $t \approx 711M$, the time identified with a blue dot. (This time results from the parameter choice $\beta_{\rm TP} = 0.35$, as defined in Sec.\ \ref{sec:endoftran}.) The plunging trajectory follows the blue dot-dashed curve. During the plunge, the infalling secondary freezes by $t \approx 722M$ to the horizon coordinate $r_{\rm H} = 1.714M$; the inset panel zooms in on the radial trajectory as the secondary approaches and freezes to the horizon. The axial motion (bottom panel) shows small oscillations as $\phi$ accumulates, at least until the motion freezes in these coordinates at the horizon.  After reaching the horizon, $\phi$ evolves linearly with a slope of $\Omega_{\rm H} = a/(2Mr_{\rm H}) = 0.204/M$.}
    \label{fig:WL_e0_0.1_a_0.7}
\end{figure}

Figure \ref{fig:WL_e0_0.1_a_0.7} displays an example of a prograde worldline constructed using our inspiral-transition-plunge prescription.  The top panel shows the radial motion versus Boyer-Lindquist time $t$ for a small body spiraling into a Kerr black hole with spin parameter $a = 0.7M$ at a mass ratio of $\eta = 10^{-4}$. At the onset of the transition, the orbital eccentricity is $e_{\rm IT} = 0.501$.  The bottom panel of Fig. \ref{fig:WL_e0_0.1_a_0.7} plots the axial coordinate $\phi$ versus $t$.

The radial motion presented in the top panel of Fig.\ \ref{fig:WL_e0_0.1_a_0.7} includes the final few cycles of adiabatic inspiral, shown as the dashed black curve.  The red dot at $t \approx 594M$ denotes when adiabatic inspiral has ended, following the prescription discussed in Sec.\ \ref{sec:endofinspiral} and using $\alpha_{\rm IT} = 0.8$; the solid red curve traces out the transition trajectory that follows.  The transition in this case executes about three-fourths of a radial period, reaching the LSO at $t \approx 708M$, a moment marked by a black bar on the trajectory.  The blue dot at $t \approx 711M$ designates when the transition ends, following the prescription discussed in Sec.\ \ref{sec:endoftran} and using $\beta_{\rm TP} = 0.35$; the dot-dashed blue curve which follows describes the final plunging geodesic.  Using Boyer-Lindquist time, the plunging secondary freezes at the horizon by $t \approx 722M$.  (We explore how the inspiral-transition-plunge worldline depends on $\alpha_{\rm IT}$ and $\beta_{\rm TP}$ in the following section.)

The axial motion shown in the bottom panel of Fig.\ \ref{fig:WL_e0_0.1_a_0.7} exhibits small amplitude oscillations about secular growth throughout the inspiral and transition.  As the secondary approaches the event horizon, $d\phi/dt\rightarrow \Omega_{\rm H}$, where $\Omega_{\rm H}$ is the angular velocity of the horizon defined in Eq.\ (\ref{eq:OmegaH}), here taking the value $\Omega_{\rm H} = 0.204/M$.  As parameterized by Boyer-Lindquist time, the small body freezes on the rotating horizon, ``whirling'' such that $\phi$ grows linearly at the rate $\Omega_{\rm H}$.

\subsection{Robustness of our worldlines versus initial conditions and model parameter choices}
\label{subsec:ResultsB}

In both the transition and plunge regimes, our worldlines show a dependence on the trajectory's initial radial anomaly angle $\chi_{r0}$, the parameter $\alpha_{\rm IT}$ that sets the end of adiabatic inspiral, and the parameter $\beta_{\rm TP}$ that sets the end of the transition.  Here we investigate the sensitivity of our worldlines to these parameters.  As we describe in detail below, the initial radial phase has an appreciable impact on a system's behavior as it enters and evolves through the transition and plunge.  This is an important physical effect, reflecting the fact that systems which start entirely identical for except their initial radial phases may undergo quite different motion as adiabatic inspiral comes to an end.

The parameters $\alpha_{\rm IT}$ and $\beta_{\rm TP}$ are {\it ad hoc} parameters which divide the systems' evolution into regimes of inspiral, transition, and plunge.  Because they essentially allow us to switch the system from one method of approximating its dynamics to another, one would hope that they have little influence on the binary dynamics that we find.  We in fact find that the parameter $\beta_{\rm TP}$, which separates transition from plunge, has very little influence on the system's behavior across a wide range of plausible parameter values.  By contrast, we find a non-trivial dependence on the parameter $\alpha_{\rm IT}$.  We are confident that this is an artifact of how we somewhat artificially divide the system's evolution into ``inspiral'' and ``transition.''  As we discuss in our conclusions, a more extensive analysis of the transition is likely needed to develop a better model for how inspiral ends for eccentric systems.

\subsubsection{Initial radial phase $\chi_{r0}$}

\begin{figure*}[ht]
    \centering
    \includegraphics[scale = 0.55]{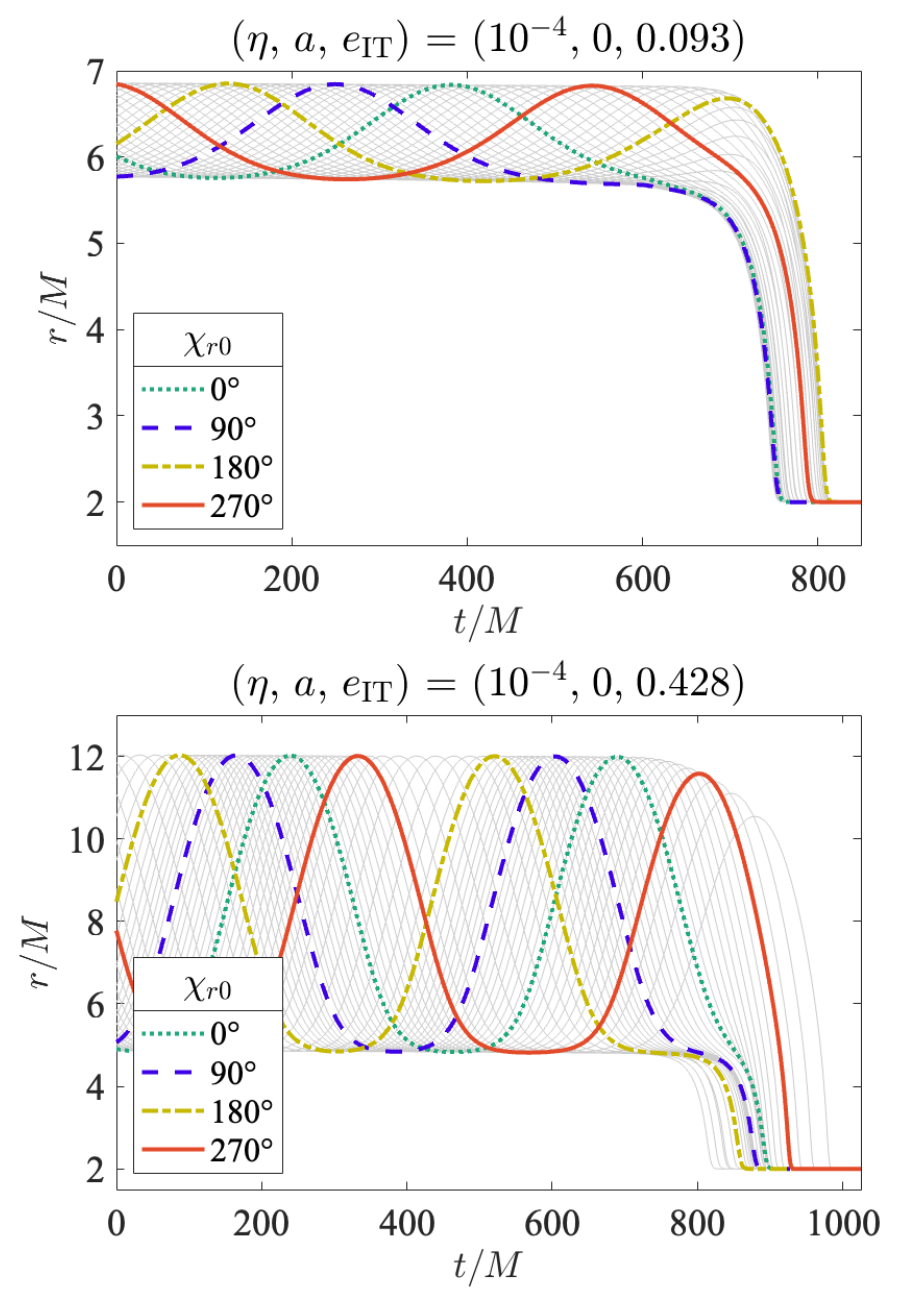}
    \includegraphics[scale = 0.495]{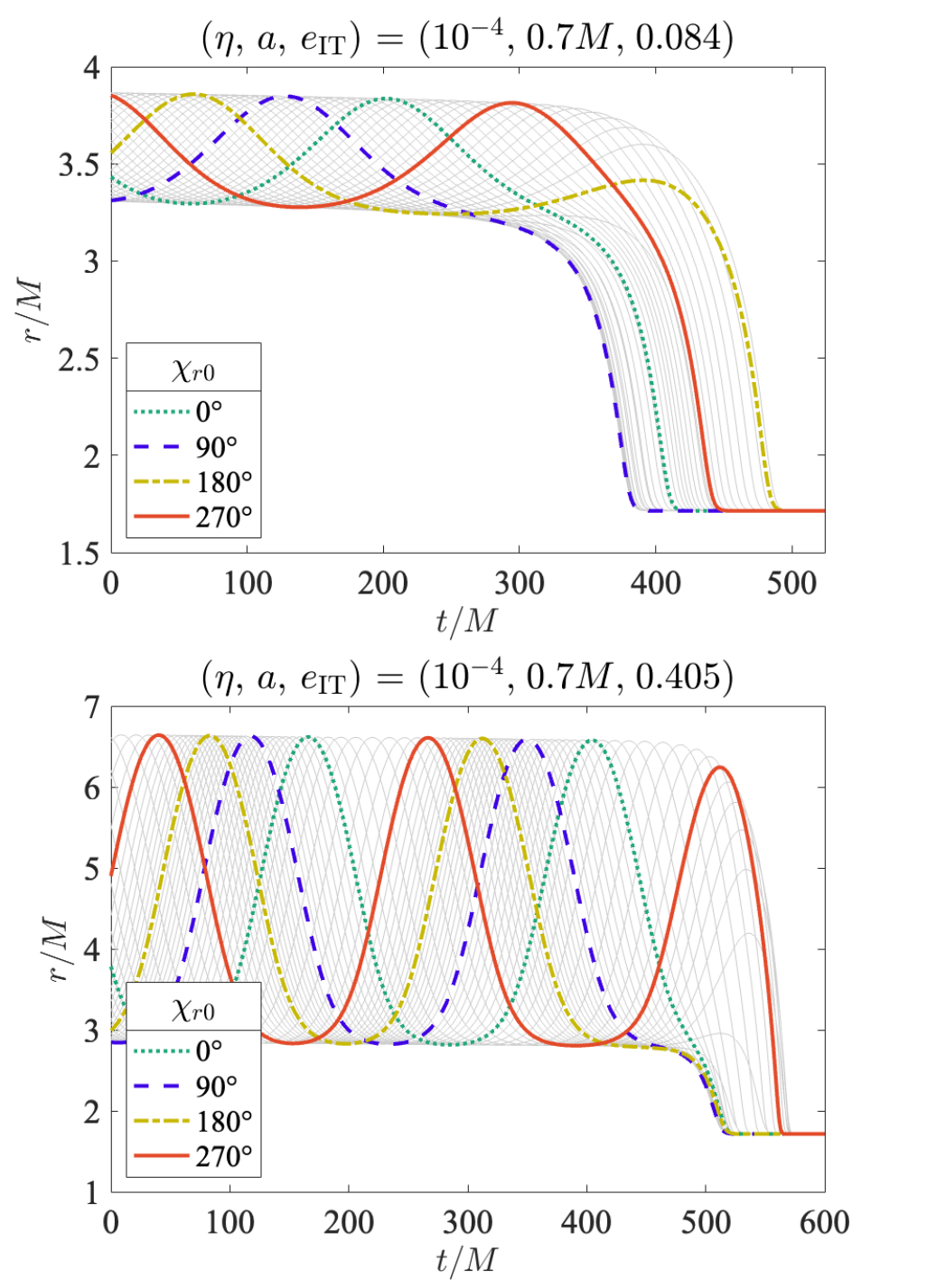}
    \caption{Prograde worldlines with different values of initial angle $\chi_{r0}$.  Left-hand panels are for inspiral, transition, and plunge into a Schwarzschild black hole; right-hand panels are into a Kerr black hole with spin parameter $a = 0.7M$.  Top panels show results for small eccentricity at the time of transition ($e_{\rm IT} = 0.093$ for Schwarzschild, $e_{\rm IT} = 0.084$ for Kerr); bottom panels are for relatively large eccentricity at transition ($e_{\rm IT} = 0.428$ for Schwarzschild, $e_{\rm IT} = 0.405$ for Kerr).  In all panels, the systems have mass ratio $\eta = 10^{-4}$, the end of inspiral parameter $\alpha_{\rm IT} = 0.8$, and the end of transition parameter $\beta_{\rm TP} = 0.35$.  In each case, we show tracks for 36 different trajectories, sampling in initial radial anomaly angle $\chi_{r0}$ uniformly from the interval $[0\degree, 360\degree]$ with step size $\Delta \chi_{r0} = 10\degree$.  We highlight the trajectories with $\chi_{r0} = 0^\circ$, $90^\circ$, $180^\circ$, $270^\circ$.  All cases show that $\chi_{r0}$ has a strong influence on the timing of the final plunge, as well as on the dynamics of the secondary's final cycle before plunge.  For some values of $\chi_{r0}$, the secondary plunges soon after it enters the transition (e.g., $\chi_{r0} = 0\degree$ and $\chi_{r0} = 90\degree$ in both the Kerr cases); for others, the small body executes an additional radial oscillation before plunging (e.g., $\chi_{r0} = 180\degree$ in both top panels, and $\chi_{r0} = 270\degree$ in both bottom panels).  Across this sampling of cases, we see significant variation in the amplitude of the final cycle of radial motion, highlighting the effect this initial condition has on the binary's final dynamics before plunge.}
    \label{fig:chir0Compare}
\end{figure*}

Figure \ref{fig:chir0Compare} examines several sets of prograde worldlines.  We show example worldlines for inspiral, transition, and plunge into Schwarzschild black holes, as well as examples going into Kerr black holes with spin $a = 0.7M$.  In all cases, the systems under consideration have mass ratio $\eta = 10^{-4}$, and transition model parameters $\alpha_{\rm IT} = 0.8$ and $\beta_{\rm TP} = 0.35$.  The top panels of Fig.\ \ref{fig:chir0Compare} study systems with small eccentricity at the start of transition ($e_{\rm IT} = 0.093$ for Schwarzschild, $e_{\rm IT} = 0.084$ for Kerr); the bottom panels show results for systems with moderately large eccentricity at the start of transition ($e_{\rm IT} = 0.428$ for Schwarzschild, $e_{\rm IT} = 0.405$ for Kerr).  All panels include 36 trajectories, differing by their initial\footnote{Note that ``initial'' here refers to the value we choose to set up the adiabatic inspiral.  Because we re-align the origin of our time coordinate in order to construct the inspiral-transition-plunge worldline, the value of $\chi_{r0}$ does not correspond to the $t = 0$ radial behavior shown in this figure.  There is, however, a one-to-one correspondence between the value of $\chi_{r0}$ and the dynamics we show, so we continue to use this label even if it no longer quite defines ``initial'' behavior in this parameterization.} mean anomaly angle $\chi_{r0}$.  The values of $\chi_{r0}$ are sampled uniformly from the range $[0\degree, 360\degree]$ with a step size of $\Delta\chi_{r0} = 10\degree$.  We highlight the trajectories with $\chi_{r0}\in \{0\degree, 90\degree, 180\degree, 270\degree\}$, showing the other 32 trajectories with lighter gray curves.  Note that we have looked at other values of spin parameter $a$.  The results we find in all cases are very similar to what we show for $a = 0$ and $a = 0.7M$, so we do not show them explicitly.

During adiabatic inspiral, the dynamics of these different trajectories are the same, albeit out of phase.  The Schwarzschild systems cross into the transition domain for $\alpha_{\rm IT} = 0.8$ at $t \approx 607M$ for the small eccentricity case, and $t \approx 644M$ for large; for the Kerr cases, the corresponding numbers are $t\approx 225M$ for small eccentricity and and $t\approx 430M$ for large.  The systems' behaviors diversify at this point.  Consider the $\chi_{r0} = 90\degree$ and $\chi_{r0} = 180\degree$ trajectories in the top-right panel (blue dashed line and yellow dashed-dotted line, respectively).  When transition begins, the $\chi_{r0} = 90\degree$ secondary plunges almost immediately. Following the light gray contours, we see that as we increase $\chi_{r0}$, the pre-plunge trajectory gradually elongates, delaying the final plunge.  Eventually, the plunge has been stalled so much that the small body completes an additional radial cycle before plunging, as is the case for the $\chi_{r0} = 180\degree$ trajectory in the top-right panel.  The light gray contours also show that when there is an additional radial revolution in the transition, the amplitude of this cycle is influenced by $\chi_{r0}$.  Similar behaviors can be seen in the other cases presented in this figure.
 
These variations in the amplitude of the final cycle and the timing of the plunge are tied to the range of initial conditions that a binary can have as it enters the transition: trajectories that move into the transition with sizable, inward-directed velocities typically plunge without an additional cycle (e.g., the $\chi_{r0} = 90\degree$ trajectory in all panels), while those that enter transition with outgoing velocities tend to continue moving outward and turn around before plunging (e.g., the $\chi_{r0} = 180\degree$ trajectory in both top panels, and the $\chi_{r0} = 270 \degree$ trajectory in both bottom panels).  The secondary's final path to plunge is thus strongly influenced by the initial anomaly angle, since it determines a binary's conditions as it enters the transition.  Two binary systems can be identical in all characteristics but this angle; but, because of this different anomaly, their final dynamics may vary greatly, leading to differences in their last orbital oscillations and the timing of their final plunge.  In our conclusions, we discuss how we expect this dependence will affect the GWs produced by these systems.

\subsubsection{From inspiral to transition: The parameter $\alpha_{\rm IT}$}
\label{subsec:insptotran}

We now examine how the worldlines we construct depend on our choice of $\alpha_{\rm IT}$, the parameter that defines end of inspiral and beginning of transition.  As discussed in Sec.\ \ref{sec:endofinspiral}, our diagnostic for the end of adiabatic inspiral uses a ratio of two accelerations: the generally inward-directed\footnote{The direction associated with this acceleration may reverse sign as the LSO is approached.} acceleration of periapse due to the backreaction of gravitational radiation, and the outward-directed geodesic coordinate acceleration at periapse.  When this ratio exceeds $\alpha_{\rm IT}$, we end the inspiral and describe the system as in the transition to plunge. 

\begin{figure*}[ht]
    \includegraphics[scale = 0.525]{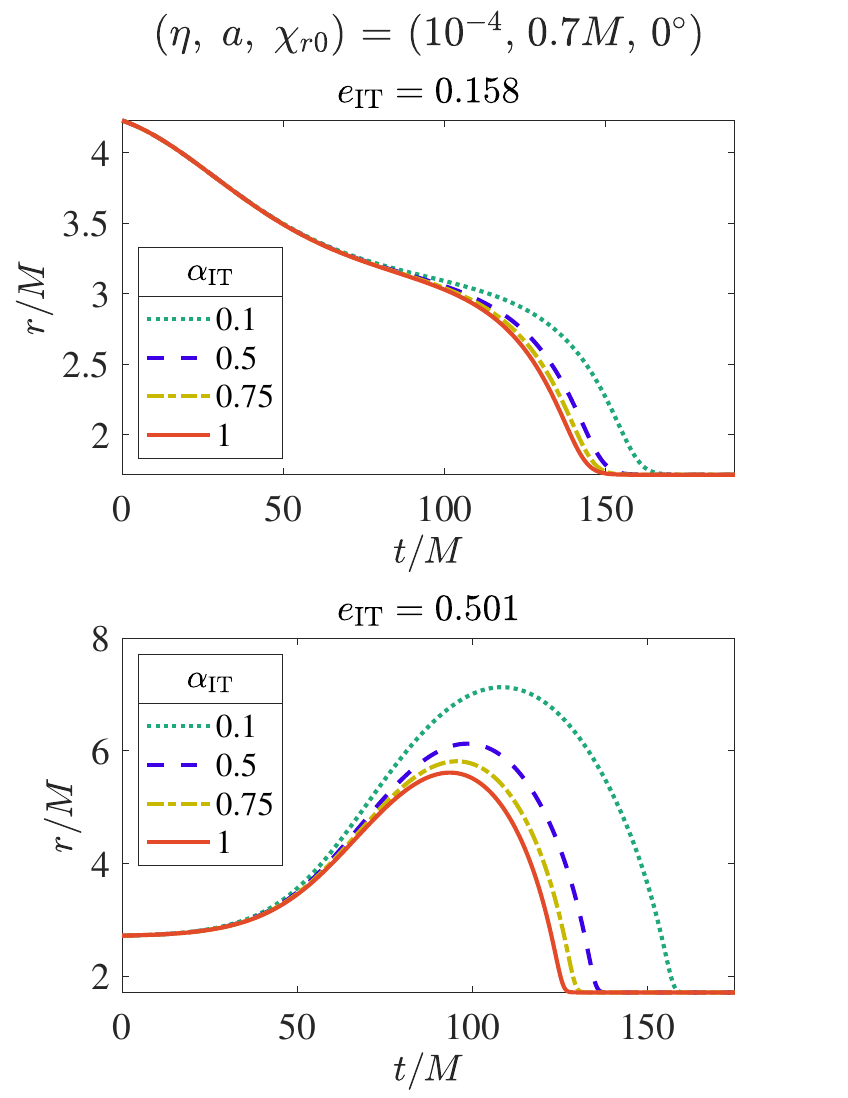}
    \hfill
    \includegraphics[scale = 0.435]{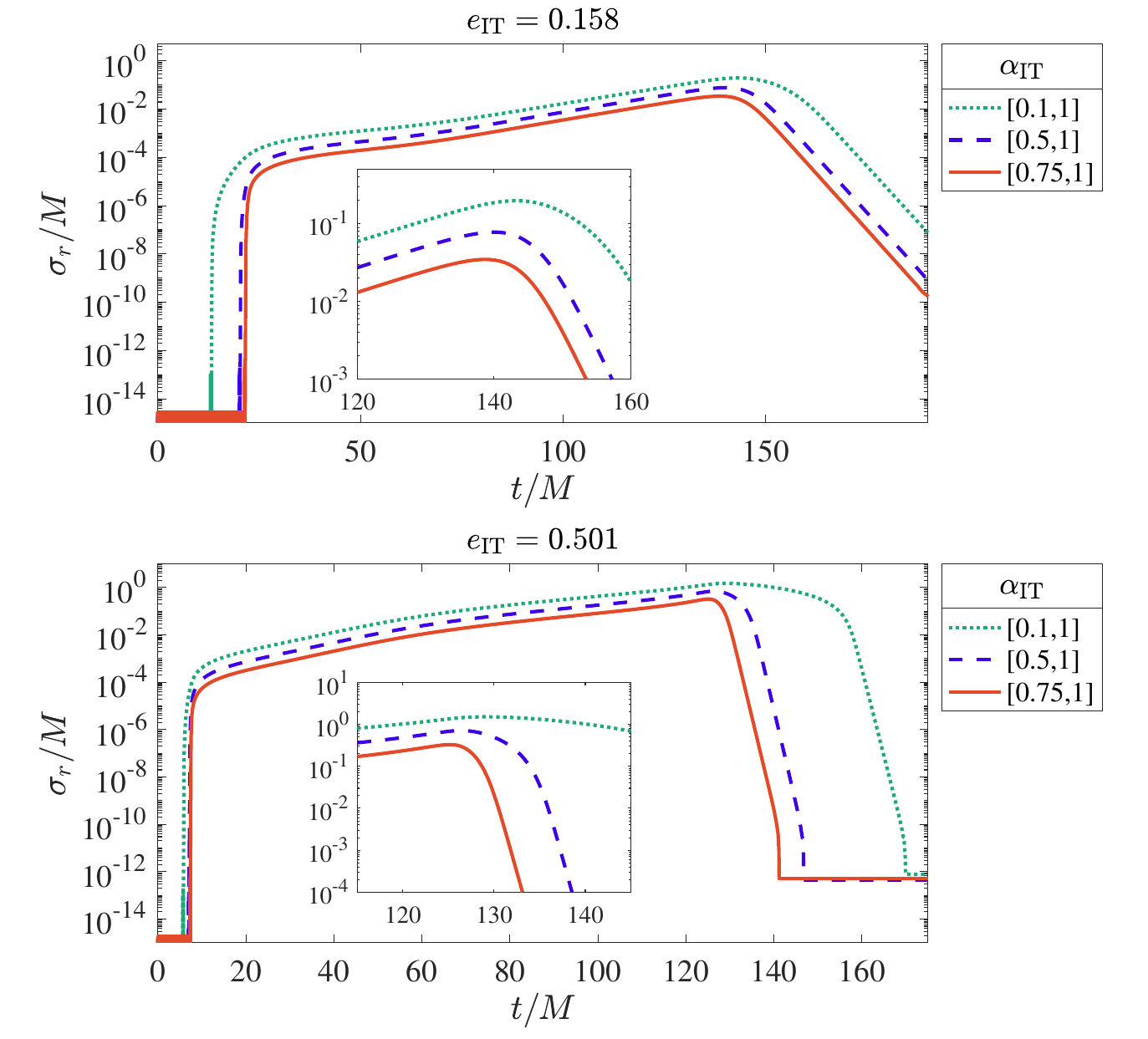}
    \caption{Left panels: a study of how prograde trajectories vary with the parameter $\alpha_{\rm IT}$.  Right panels: standard deviation, as defined by Eq.\ (\ref{eq:std}), of the distribution with respect to $\alpha_{\rm IT}$ of a trajectory's radial position versus Boyer-Lindquist coordinate time.  In all panels, the binaries under study have parameters $\eta = 10^{-4}$, $a = 0.7M$, $\chi_{r0} = 0\degree$, and $\beta_{\rm TP} = 0.35$; $t = 0$ in these plots marks the onset of the transition for the $\alpha_{\rm IT} = 0.1$ case.  (The transition begins slightly later as $\alpha_{\rm IT}$ is increased.)  The top panels show results for binaries for which the eccentricity $e_{\rm{IT}} = 0.158$ when transition begins; for the bottom panels, the binaries have $e_{\rm{IT}} = 0.501$.  The standard deviations in the right panels are calculated from $N = 91$ worldlines across different ranges of $\alpha_{\rm IT}$, as indicated by the inset to the right of each set of curves.  The standard deviations traced out by the green, dotted lines are calculated using trajectories spanning the range $\alpha_{\rm IT} \in [0.1, 1]$; while the other curves study focus on reduced ranges of $\alpha_{\rm IT}$. The inset below each set of curves zooms in on the region surrounding the largest standard deviations.}
    \label{fig:alphaIT}
\end{figure*}

In Figure \ref{fig:alphaIT}, we investigate the impact of choosing $\alpha_{\rm IT}$ from the range $[0.1, 1]$ for inspiral, transition, and plunge into a Kerr black hole with $a = 0.7M$.  Each panel in this figure describes a binary with $\eta = 10^{-4}$, with an initial mean anomaly of $\chi_{r0} = 0\degree$.  Time $t = 0$ labels the beginning of the transition for the $\alpha_{\rm IT} = 0.1$ case; larger values of $\alpha_{\rm IT}$ lead to a later start for the transition.  Note that for $t < 0$, each case follows an identical inspiral, so we omit that portion of the trajectory.  All data included in the top panels of this figure correspond to trajectories for which the eccentricity at transition\footnote{Because changing $\alpha_{\rm IT}$ changes the time at which transition begins, the value of $e_{\rm IT}$ varies slightly across trajectories within each panel.  For the range of $\alpha_{\rm IT}$ shown here, we find variations $\delta e_{\rm IT} \lesssim 10^{-3}$.  We use the value of $e_{\rm IT}$ for $\alpha_{\rm IT} = 0.1$ to label our plots.} is $e_{\rm IT} \approx 0.158$; those in the bottom panels have $e_{\rm IT} \approx 0.501$.

The left-hand panels of Fig.\ \ref{fig:alphaIT} show $r$ versus $t$ for several $\alpha_{\rm IT}$ values.  These two panels illustrate the two types of behavior we observe as we sample $\alpha_{\rm IT}$ over the range $[0.1, 1]$: an offset in the time at which plunge begins (which leads to dephasing in the motion), as seen in the smaller eccentricity case in the top left; and a spread in the amplitude of the final radial cycle, as seen in the larger eccentricity case in the bottom left.  (This spread in the amplitude also involves an offset in the time of plunge.)  The right-hand panels of Fig.\ \ref{fig:alphaIT} examine the distribution of behaviors we see as $\alpha_{\rm IT}$ is varied.  We build $N$ trajectories for $N$ choices of $\alpha_{\rm IT}$, and define the ``trajectory radial standard deviation'' $\sigma_r(t)$ as
\begin{equation}
    \sigma_r(t) = \sqrt{\f{1}{(N-1)} \sum_{i = 1}^N |r_i(t) - \mu_{r}(t)|^2}\;.
\label{eq:std}
\end{equation}
Here, $r_i(t)$ is the radial trajectory generated by the $i$th choice of $\alpha_{\rm IT}$, and
\begin{equation}
    \mu_r(t) = \f{1}{N} \sum_{i = 1}^N r_i(t)
\label{eq:mean}
\end{equation}
is the mean radial trajectory found for the sample.

For the cases displayed in the right-hand panels of Fig.\ \ref{fig:alphaIT}, we use $N = 91$, selecting $\alpha_{\rm IT}$ uniformly from the interval indicated by the key in the figure.  When we take $\alpha_{\rm IT}$ from the ``full'' range $[0.1, 1]$ that we consider (dotted green curves), we find the greatest $\sigma_r(t)$ in all cases, peaking at $\sigma_r(t) \simeq 0.198M$ for our small eccentricity case, and at $\sigma_r(t) \simeq 1.51M$ for the larger eccentricity case. This is consistent with the significant spread in radial trajectories in the left-hand panels of this figure.

Much of this spread in radial trajectories is due to the inclusion of small values of $\alpha_{\rm IT}$: smaller values of $\alpha_{\rm IT}$ lead to changing from inspiral to transition at early times.  The right-hand panels of Fig.\ \ref{fig:alphaIT} also examine what happens when we take our $N = 91$ samples from the range $[0.5, 1]$ and from $[0.75, 1]$.  The maximum in $\sigma_r(t)$ is reduced quite a bit doing this: for the small eccentricity cases, $\sigma_r(t)$ peaks at about $0.078M$ for $\alpha_{\rm IT} \in [0.5, 1]$, and at about $0.035M$ for the range $[0.75, 1]$; for the large eccentricity cases, $\sigma_r(t)$ peaks at about $0.709M$ for $\alpha_{\rm IT} \in [0.5, 1]$, and about $0.328M$ for the range $[0.75, 1]$.

Examining the impact of $\alpha_{\rm IT}$ for other black hole spins yields similar results, so we omit figures presenting those cases in detail. For example, Schwarzschild trajectories with a small transition eccentricity of $e_{\rm IT} = 0.171$ have $\sigma_r(t)$ that peaks at about $0.845M$ for $\alpha_{\rm IT} \in [0.1, 1]$.  This reduces to $0.426M$ for $\alpha_{\rm IT} \in [0.5, 1]$ and to $0.208M$ for $\alpha_{\rm IT} \in [0.75, 1]$.  Numbers for high eccentricity trajectories are similar.  The Schwarzschild case with $e_{\rm IT} = 0.524$ has a maximum $\sigma_r(t)$ of about $0.857M$ when $\alpha_{\rm IT} \in [0.1, 1]$ that decreases to about $0.496M$ for $\alpha_{\rm IT} \in [0.5, 1]$ and to about $0.246M$ when $\alpha_{\rm IT} \in [0.75, 1]$.

In all cases, we find that smaller values of $\alpha_{\rm IT}$ produce greater variations in our worldlines than when $\alpha_{\rm IT}$ is close to unity.  In order to minimize the influence of this {\it ad hoc} parameter, we typically use $\alpha_{\rm IT} = 0.8$ in our studies.  However, it must be noted that we do not have a strong physical motivation for this choice.  A first-principles analysis of these systems which carefully studies the breakdown of adiabaticity and the end of inspiral, akin to that done in Ref.\ \cite{kuchler2024} for configurations with zero eccentricity, would significantly improve our understanding of this behavior.

\subsubsection{From transition to plunge: The parameter $\beta_{\rm TP}$}
\label{subsec:trantoplunge}

We conclude this section by examining how our worldlines depend upon the parameter $\beta_{\rm TP}$.  As described at length in Sec.\ \ref{sec:endoftran}, this parameter controls when we end the transition and change to a plunging geodesic.  When $\beta_{\rm TP} = 0$, our plunge begins exactly at the LSO, making the ``transition'' part of the worldline negligible; when $\beta_{\rm TP} = 1$, the transition goes all the way to the event horizon, making the ``plunge'' negligible.

\begin{figure*}[ht]
    \includegraphics[scale = 0.48]{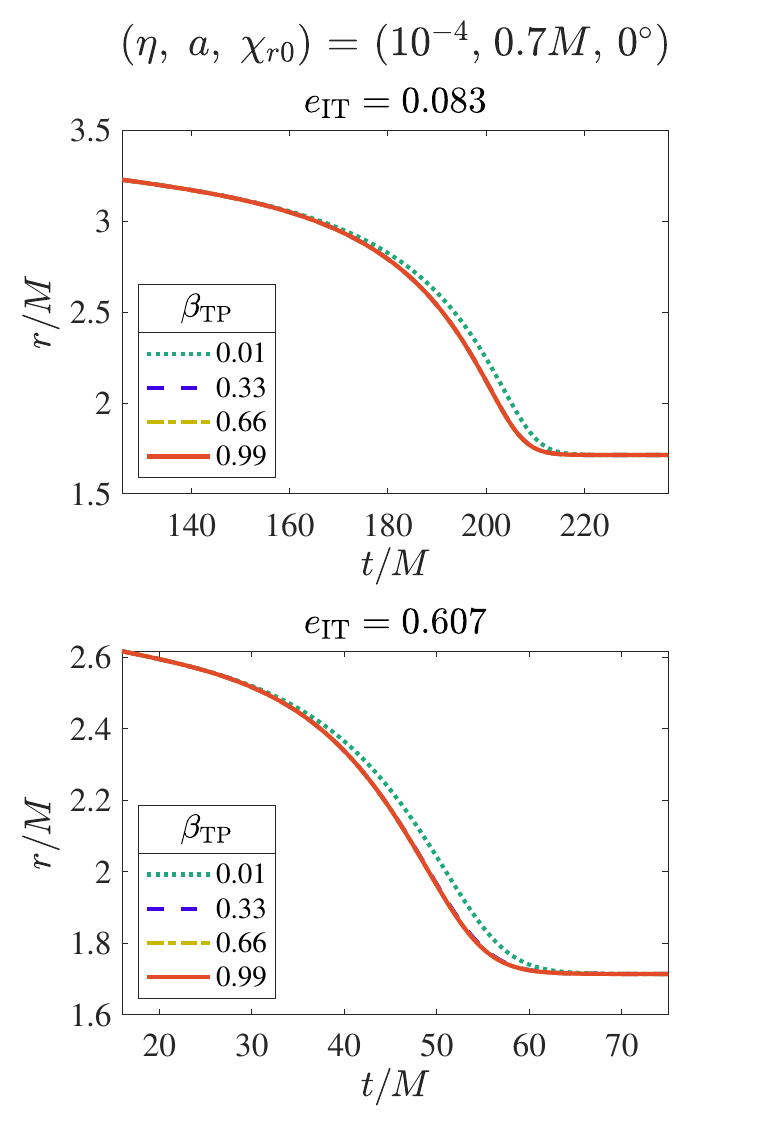}
    \hfill
    \includegraphics[scale = 0.51]{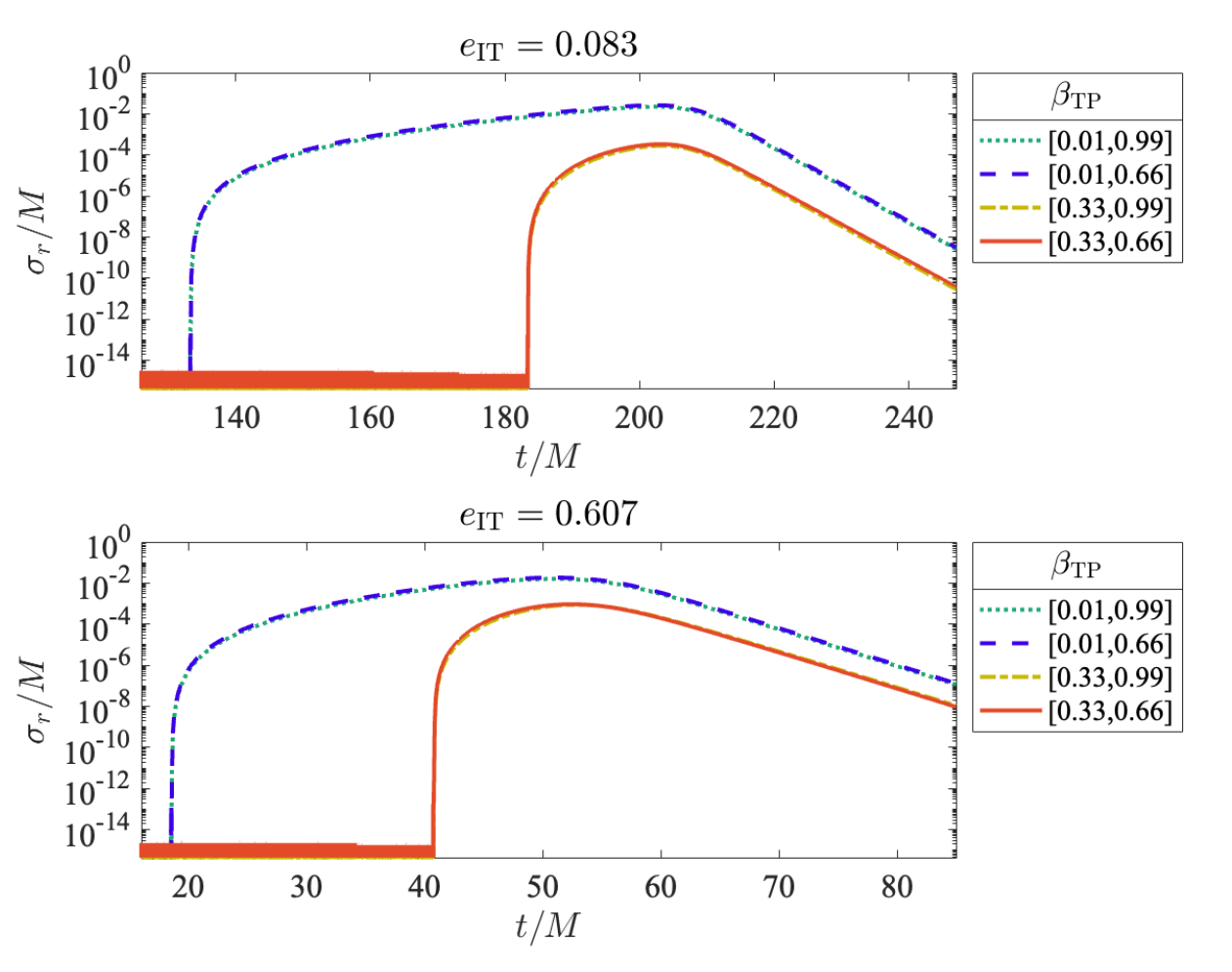}
    \caption{Left panels: a study of how prograde trajectories vary with the parameter $\beta_{\rm TP}$.  Right panels: standard deviation, as defined by Eq.\ (\ref{eq:std}), of the distribution with respect to $\beta_{\rm TP}$ of a trajectory's radial position versus Boyer-Lindquist coordinate time.  In all panels, the binaries under study have parameters $\eta = 10^{-4}$, $a = 0.7M$, $\chi_{r0} = 0\degree$, and $\alpha_{\rm IT} = 0.8$.  The top panels show results for binaries with eccentricity $e_{\rm IT} = 0.083$ when transition begins; for the bottom panels, the binaries have $e_{\rm IT} = 0.607$.  Notice in the left-hand panels that, with the exception of the choice $\beta_{\rm TP} = 0.01$, the trajectories cannot be clearly distinguished at this resolution.  The right-hand panels show $\sigma_r(t)$ using $N = 99$ trajectories, varying $\beta_{\rm TP}$ over the range indicated by the inset panel.  In all cases, we find the maximum standard deviation in this sample $\sigma^{\rm max}_r(t) \sim 10^{-2}M$ when our range of trajectories has a lower bound on $\beta_{\rm TP}$ of $0.01$.  When our range excludes trajectories with such small values of $\beta_{\rm TP}$, the spread in our distribution shrinks to $\sigma^{\rm max}_r(t) \sim 10^{-4}M$.}
    \label{fig:blso_compare}
\end{figure*}

Figure \ref{fig:blso_compare} shows how changing $\beta_{\rm TP}$ affects the worldlines we construct.  In all panels, we present the radial behavior of worldlines for systems with $\eta = 10^{-4}$, $a = 0.7M$, $\chi_{r0} = 0\degree$, and $\alpha_{\rm IT} = 0.8$.  The top panels include results for trajectories whose eccentricity at the start of the transition is $e_{\rm IT} = 0.083$; in the bottom panels, $e_{\rm IT} = 0.607$.  In the left-hand panels, we examine $r$ versus $t$ for a number of choices of $\beta_{\rm TP}$ (only showing data from the LSO onward, since the trajectories are identical up to that point).  With the exception of the parameter choice $\beta_{\rm TP} = 0.01$, the radial trajectories are practically indistinguishable from each other at both values of eccentricity.  This finding is confirmed in the right-hand panels of Fig.\ \ref{fig:blso_compare}.  Here we show $\sigma_r(t)$ calculated with $N = 99$ trajectories  assembled by uniformly sampling $\beta_{\rm TP}$ from $\beta_{\rm TP} \in [\beta^{\rm min}_{\rm TP}, \beta^{\rm max}_{\rm TP}]$.  When $\beta^{\rm min}_{\rm TP} = 0.01$, $\sigma_r$ can be as large as about $0.01M$ for both eccentricities.  Narrowing the distribution by choosing $\beta^{\rm min}_{\rm TP} = 0.33$ reduces this spread significantly; the maximum deviation we find is $\sigma_r(t) \sim 10^{-4}M$ in this case.  Interestingly, adjusting $\beta^{\rm max}_{\rm TP}$ has little impact on this result: $\sigma_r(t)$ constructed from distributions with $\beta^{\rm max}_{\rm TP} = 0.66$ can barely be distinguished from $\sigma_r(t)$ constructed using $\beta^{\rm max}_{\rm TP} = 0.99$.  This appears to reflect the fact that the transition trajectory is very similar to the plunge once the system has evolved by about $\delta r \sim M$ beyond the LSO.

Much like what we found for $\alpha_{\rm IT}$, we find similar results when we examine the impact of $\beta_{\rm TP}$ for different black hole spins.  When $a = 0$, the values of $\sigma_r(t)$ decrease by orders of magnitude when we shift $\beta^{\rm min}_{\rm TP}$ from 0.01 to 0.33. For instance, the low eccentricity case with $e_{\rm IT} = 0.092$ has a maximum spread of $\sigma_r(t)\sim 0.1M$ when $\beta^{\rm min}_{\rm TP} = 0.01$.  This shrinks to $\sigma_r(t)\sim 10^{-3}M$ when $\beta^{\rm min}_{\rm TP} = 0.33$.  Changing the maximum value in the range of $\beta_{\rm TP}$ has negligible effect.  In a higher eccentricity case with $e_{\rm IT} = 0.627$, we see a fall from $\sigma_r(t)\sim 10^{-2}M$ to $\sigma_r(t)\sim 10^{-4}M$ when $\beta^{\rm min}_{\rm TP}$ is increased from $0.01$ to $0.33$; changing $\beta^{\rm max}_{\rm TP}$ has very little impact on these results.

The worldlines we construct thus exhibit a rather weak dependence on the choice of $\beta_{\rm TP}$, independent of the eccentricity and the black hole spin.  Any choice of $\beta_{\rm TP}$ from the range $[0.01, 0.99]$ yields largely the same overall plunge dynamics; the slight deviations we find can be reduced even further by narrowing that range to $[0.33, 0.99]$.  Planned work will examine the impact of this parameter on the GW spectrum produced by such systems, and it will be an interesting exercise to ascertain how much systematic effect is introduced into waveforms with this parameter.  As we concluded in our discussion of the parameter $\alpha_{\rm IT}$, a first-principles analysis may also help us to better understand the limitations of this framework.  It is reassuring, though, that this {\it ad hoc} parameter appears to have only a rather minor impact on the dynamics of these systems.

\section{Conclusions}
\label{sec:conclude}

In this paper, we have constructed tools to determine the end of a small-mass-ratio eccentric binary's adiabatic inspiral, and to smoothly connect the worldline describing such an inspiraling system to the final plunge.  Our approach builds a transition solution connecting inspiral to plunge that follows the conceptual framework of the pioneering model developed in OT00 \cite{OriThorne2000}.  As in OT00, our model uses input from frequency-domain black hole perturbation theory to compute how an orbit's integrals of motion evolve due to the backreaction of GWs.  Our description of the smaller body's motion through the transition to plunge is based on using the radial geodesic equation with the orbit integrals evolving due to backreaction, doing so in a regime in which the change is too rapid to be described by an adiabatic approximation.

The analysis of OT00 (which focuses on circular and equatorial orbits) demonstrated that the radial behavior through the transition and plunge can be described using a fairly simple ``universal'' solution which, by using simple scaling rules, describes the post-inspiral radial behavior for all\footnote{At least all mass ratios small enough that it is valid to use black hole perturbation theory.} black hole spins and mass ratios.  The model developed in AH19 \cite{ApteHughes2019} showed that this universal radial solution works for all quasi-circular configurations, including those for which the orbits are inclined out of the equatorial plane.

In strong contrast to these previous works, the solution that we find for transition and plunge following eccentric equatorial inspiral does {\it not} admit a universal solution for the radial behavior.  We instead find radial dynamics which depend very strongly on the phase of the radial motion.  That the transition and plunge will have such a strong dependence on this phase is clear on physical grounds: two adiabatic inspirals which are identical except for the starting value of this phase may enter the transition in very different radial configurations.  One may approach the transition near periapsis, and proceed quickly to plunge; the other may enter the transition near apoapsis and require an additional half cycle of radial motion before plunging.  This dependence on the radial phase leads to a substantial spread in the final behavior of otherwise identical eccentric inspiraling systems, as shown in Fig.\ \ref{fig:chir0Compare}.  We note that we find this behavior for all the cases that we have examined, which includes both additional spin values for prograde inspiral and plunge and several examples of retrograde\footnote{Because the retrograde LSO is at larger values of $p$ than the prograde LSO, the GW-driven inspiral of the periapsis tends to be significantly smaller than in the prograde case.  For large values of $a$, it may thus be necessary for retrograde adiabatic inspiral data to come even closer to the LSO than the offset $\delta p = 10^{-4}M$ that, as discussed in Sec.\ \ref{sec:adiabinsp}, we used throughout this analysis.} inspiral and plunge.

Much like the models introduced in OT00 and AH19, our framework requires a pair of {\it ad hoc} parameters which determine when we switch our description from inspiral to transition, and when we switch from transition to plunge.  We switch from inspiral to transition by examining the rate at which the changing location of the orbit's periapsis accelerates due to GW-driven backreaction, and comparing that with the ``geodesic'' acceleration at periapsis in the absence of GW emission.  When the ratio of these accelerations exceeds a threshold, $a_{\rm p,GW}/a_{\rm p,geod} \ge \alpha_{\rm IT}$, the system is no longer evolving adiabatically.  We declare inspiral to have ended at this point, and switch to our transition-regime equations of motion.  We then switch from transition to plunge at a radius that is some fraction of the coordinate distance between the location of periapsis on the LSO and the event horizon.  We parameterize this fraction with a quantity $\beta_{\rm TP}$ whose value we chose between $0$ (denoting a plunge that begins at the LSO) and $1$ (denoting a transition that reaches all the way to the event horizon).

The parameters $\alpha_{\rm IT}$ and $\beta_{\rm TP}$ play roles similar to the parameters $L_i$ and $L_f$ which, in AH19, bounded the initial and final moments of the universal transition solution used in that analysis.  We do not have a physical model for $(\alpha_{\rm IT}, \beta_{\rm TP})$, but instead examine the behavior of the worldlines that we produce as we vary these parameters over a plausible range.  We find very little dependence on the parameter $\beta_{\rm TP}$, especially if we confine this parameter to the domain $\beta_{\rm TP} \ge 0.33$.  This seems to indicate that the transition solution is not very different from a plunging geodesic over much of the domain that we cover.  We find a much larger dependence on the parameter $\alpha_{\rm IT}$, though this dependence is reduced if we restrict to the domain $\alpha_{\rm IT} \ge 0.5$.  For zero eccentricity binaries, more detailed and complete physical models describing the end of adiabatic inspiral and the transition to plunge have recently become available \cite{CompereKuchler2021, kuchler2024}.  Such models can help us calibrate appropriate choices for our arbitrary parameters, or do away with them altogether.  Our analysis indicates that extending such models to include eccentricity could be particularly important, as ambiguity in precisely defining the end of inspiral can lead to significant variation in the worldline that a model predicts.

Accepting the limitations introduced by these ambiguities, we plan to use this framework to study how eccentricity influences the late gravitational waveforms produced by binary systems.  To generate waveforms, we will mirror the procedure used in Ref.\ \cite{Lim2019}: following the prescription described in this paper, we will build inspiral, transition, and plunge worldlines followed by a small body moving on an initially eccentric orbit, and use them to develop the source-term of the time-domain Teukolsky equation \cite{Teukolsky1973, Sundararajan2007, Sundararajan2008, Sundararajan2010, Zenginoglu2011}.  Our goal is to understand to what extent a binary's final eccentricity ``colors'' the late-time gravitational waveforms that it produces.

Based on the ``WKB intuition'' that a system's late ringdown waves can be understood as wavepackets gradually leaking out of a black hole's light ring (see, for example, Refs.\ \cite{FerrariMashhoon1984, SchutzWill1985} for discussion), we speculate that the ringdown structure of eccentric inspirals and plunges may be qualitatively rather different from the ringdown structure of quasi-circular inspirals and plunges.  The orbital geometry of eccentric systems opens the possibility of wavepackets being excited into, and then slowly leaking from, both a black hole's retrograde and prograde light rings in rapid succession\footnote{This intuitive picture arose from a discussion of these systems with Yanbei Chen; we thank him for the free-ranging discussion in which this idea was developed.}.  Preliminary waveforms developed by G.\ Khanna as part of our planned follow-up analysis indeed indicate that eccentric systems produce a more complicated ringdown spectrum than we find in the quasi-circular case.  As part of this follow-up analysis, we aim to systematically examine how this more complicated spectrum can be understood as a superposition of different ringdown modes, and how the excitation of these modes depends on system parameters like the eccentricity and the spin of the primary.  Understanding how eccentricity is imprinted upon the late mode structure may make it possible to constrain the geometry of these systems from their late-time coalescence waveforms.  Seeking such an understanding was a major motivator of past work studying misaligned quasi-circular systems \cite{Hughes2019, ApteHughes2019, Lim2019}.  It is very encouraging that analyses of recent GW events (e.g., Ref.\ \cite{Siegel2023}) is consistent with the suggestions of these past analyses.

The effective-one-body (EOB) and numerical relativity (NR) communities are also investigating signatures of eccentricity on late-time waveforms (see, e.g.,\ Refs.\ \cite{EOB_ecctran, Carullo_2024} for an EOB description of the eccentric transition to plunge in the nonspinning, small-mass-ratio limit, Ref.\ \cite{Sperhake_2018} for the transition to plunge from eccentric NR simulations in the non-spinning, equal-mass binary regime, and Refs.\ \cite{ficarralousto2024, wang2024_1, wang2024_2} for recent studies of eccentric binary black hole coalescences with NR).  We expect it will be quite fruitful to compare the waveforms produced by our models with those from EOB and NR.  The EOB comparison will allow us to compare and assess our criterion for evaluating the end of inspiral, perhaps providing some insight into calibration of parameters $\alpha_{\rm IT}$ and $\beta_{\rm TP}$.  Comparison to NR studies will allow us to investigate the extent to which our findings in the small-mass-ratio limit extend to systems with more comparable mass ratios.

We also plan to extend the small-mass-ratio binary configurations that we consider to fully generic --- inclined and eccentric --- orbital geometries.  Such binaries are described in the adiabatic limit by the evolution of all three orbit integrals: energy $E$, axial angular momentum $L_z$, and Carter constant $Q$.  Although we specialized our detailed analysis in this paper to equatorial binaries for which $Q = 0$, the framework we have developed is general.  Upgrading the discussion of, for example, Secs.\ \ref{sec:adiabinsp} and \ref{sec:trans} to include non-zero $Q$ is straightforward, as is generalizing Sec.\ \ref{sec:worldline} to include motion in the $\theta$ direction.  Indeed, much of this was done already in AH19; making generic inspiral, transition, and plunge trajectories will largely be an exercise in fitting certain parts of AH19 into the framework presented here.  The major bottleneck is that, at present, we do not yet have large datasets describing the parameter space of generic adiabatic inspirals.  No issue of principle prevents us from constructing such datasets; the challenge is largely one of computational cost.  Developing such datasets is planned for the Fast EMRI Waveforms project, and will be prioritized as that project fills out the Kerr parameter space.

Generic inspirals depend upon two phase parameters: the radial anomaly angle $\chi_{r0}$ discussed here, and a polar anomaly angle $\chi_{\theta0}$ which indicates a small-mass-ratio binary's initial $\theta$ position.  As shown in AH19, the anomaly angle $\chi_{\theta0}$ controls the polar angle $\theta_{\rm fin}$ at which the secondary plunges into the primary's event horizon.  Lim et al.\ \cite{Lim2019} showed that $\theta_{\rm fin}$ has a significant influence on the spectrum of modes excited in inclined coalescences; two initially circular binaries that are identical in every way but their value of $\chi_{\theta0}$ can produce gravitational waveforms with very different late-time ringdown waves.  It will almost certainly be the case that the waveforms found for the generic configurations will combine the behavior seen in the circular case with any eccentricity-dependent behavior uncovered in our planned follow-on analysis.  A detailed study of this behavior will be needed in order to understand whether, for example, mode excitation behavior associated with eccentricity may be degenerate with the mode excitation associated with inclination.  Such degeneracy may affect interpretations of gravitational waves measured from inclined, eccentric, precessing systems.  We expect the small-mass-ratio limit to continue to serve as a clean laboratory for exploring such behavior, which should prove useful as gravitational wave searches go deeper and continue to find an ever wider range of astrophysical binary sources.

\section*{Acknowledgments}

We thank Yanbei Chen for interesting and helpful discussions about how substantial eccentricity near plunge may affect the ringdown structure of binary black hole waveforms, as well as Gaurav Khanna for computing preliminary waveforms associated with our worldlines (as discussed in Sec.\ \ref{sec:conclude}).  The structure of the plots in Sec.\ \ref{sec:results} that study our worldlines' dependence on the initial mean anomaly angle $\chi_{r0}$ were inspired by Peter Nee's talk \cite{NeesTalk} at the 15th International LISA Symposium.  This work was supported by NSF Grant PHY-2110384 and PHY-2409644.  Computations used MIT Kavli Institute resources at MIT's {\tt engaging} computing cluster, as well {\tt subMIT} resources at MIT Physics.

\appendix

\section{Transition for circular inspiral}
\label{app:circtrans}

In this appendix, we review the calculation of the transition between adiabatic inspiral and plunge in the limit of zero eccentricity. Our analysis follows AH19 \cite{ApteHughes2019}, which in turn follows and generalizes OT00 \cite{OriThorne2000}.  We first sketch, somewhat heuristically, how to compute the timescales which allow one to assess whether a system is in the adiabatic inspiral, or whether transition has begun.  We focus on Schwarzschild spacetime for this heuristic discussion; the concepts generalize straightforwardly to Kerr, though various terms get more complicated.  We then discuss the equation of motion during transition, showing how a universal solution describing the motion for all parameter choices appropriate to circular orbit geometry emerges.

\subsection{Timescale analysis: When does quasi-circular adiabatic inspiral end?}

Consider an orbit that is slightly perturbed from circularity. As the system evolves through the inspiral and transition, the orbital radius can be expressed as  $r_{\rm o} = r_{\rm c} + \delta r$, where $r_c$ describes a circular orbit and $|\delta r|\ll r_c$. Since circular orbits require $(\partial R/\partial r)_{r_{\rm c}} = 0$, we can expand Eq.\ (\ref{eq:acceleqn}) in $\delta r$ as:
\begin{equation}
    \frac{d^2(\delta r)}{d\lambda^2} - \frac{1}{2}\left(\frac{\partial^2R}{\partial r^2}\right)_{r_{\rm c}}(\delta r) = 0\;.
\end{equation}
Noting that $(\partial^2R/\partial r^2)_{r_{\rm c}} \le 0$ for all circular Kerr orbits (and is only zero at the innermost orbit), we define
\begin{equation}
    \Upsilon_r = \sqrt{-\frac{1}{2}\left(\frac{\partial^2R}{\partial r^2}\right)_{r_{\rm c}}}\;.
\end{equation}
A geodesic orbit, slightly perturbed in this way, would thus exhibit simple harmonic oscillations about the circular orbit radius $r_{\rm c}$ with Mino-time frequency $\Upsilon_r$.  We can convert this to a coordinate time frequency by using the factor $\Gamma$ defined by Eq.\ (\ref{eq:Gamma}) which converts, in an orbit-averaged sense, Mino-time intervals into coordinate-time intervals:
\begin{equation}
    \Omega_r = \Upsilon_r/\Gamma\;.
\end{equation}
Note that Appendix A of AH19 performs a similar calculation, computing a frequency $\omega_r$ which is conjugate to proper time along the orbit, then converting to an observer-time frequency.  However, AH19 introduces an error by converting to coordinate time using the metric element $g_{tt}$, rather than using the orbit's $dt/d\tau$ --- their Eq.\ (A4) is incorrect.  This error does not change the most important conclusions of AH19, namely how certain aspects of the system scale with parameters during the transition, but does change certain numerical factors in that calculation by several percent.

The above equations and analysis tell us that if the orbit is displaced slightly from circularity, there will be a restoring force that pushes it back to the circular configuration on a timescale $1/\Omega_r$.  As long as any disturbances to the orbit or the potential act on a timescale slower that $1/\Omega_r$, the orbit will remain quasi-circular.  In our case, the disturbance of interest is the backreaction of GW emission on the orbit.  This backreaction changes the location of the root which defines $r_{\rm c}$, causing it to drift inward with velocity\footnote{Note that Eq.\ (A5) of AH2019 gives the reciprocal of $(dr_{\rm c}/dt)^{\rm GW}$.  This typographical error does not propagate.}
\begin{equation}
    \left(\frac{dr_{\rm c}}{dt}\right)^{\rm GW} = \frac{(dE^{\rm p}/dt)^{\rm GW}}{dE^{\rm p}/dr}\;.
\end{equation}
We have introduced here the orbit's ``physical'' energy, $E^{\rm p}$.  It is related to the specific energy used elsewhere in this paper by $E^{\rm p} = \mu E$.  The quantity $(dE^{\rm p}/dt)^{\rm GW}$ describes the rate at which the orbit's energy is lost due to gravitational wave emission.  For circular orbits,
\begin{equation}
    \left(\frac{dE^{\rm p}}{dt}\right)^{\rm GW} = -\frac{64}{5}\dot{\cal E}\eta^2\left(\frac{M}{r}\right)^5\;.
\end{equation}
The factor $\eta = \mu/M$ is the system's reduced mass ratio.  With $\dot{\cal E} = 1$, this is the quadrupole formula for GW emission from circular orbits; the factor $\dot{\cal E}$ allows us to account for strong-field corrections.  Near the Schwarzschild ISCO, $\dot{\cal E} \simeq 1.14$.

The velocity $(dr_{\rm c}/dt)^{\rm GW}$ is not constant, but accelerates.  The inward acceleration of the orbit is given by
\begin{equation}
    \left(\frac{d^2r_{\rm c}}{dt^2}\right)^{\rm GW} = \left(\frac{dr_{\rm c}}{dt}\right)^{\rm GW}\frac{d}{dr}\left(\frac{dr_{\rm c}}{dt}\right)^{\rm GW}\;.
\end{equation}
The timescale on which GWs change the location of the circular orbit is the ratio of the inward velocity to the inward acceleration:
\begin{eqnarray}
    T^{\rm GW}_r &=& \frac{(dr_{\rm c}/dt)^{\rm GW}}{(d^2r_{\rm c}/dt^2)^{\rm GW}}
    \nonumber\\
    &=& \left[\frac{d}{dr}\left(\frac{dr_{\rm c}}{dt}\right)^{\rm GW}\right]^{-1}\;.
\end{eqnarray}
At large radii, $\Omega_r T^{\rm GW}_r \gg 1$ and the orbit remains quasi-circular as it evolves.  As we approach the last stable orbit, gravitational radiation acts more quickly, while at the same time the ``restoring force'' keeping the orbit circular grows weaker.  The condition $\Omega_r T^{\rm GW}_r = 1$ is a reasonable criterion for the end of adiabatic inspiral and the beginning of transition to plunge.

For circular orbits in Schwarzschild,
\begin{eqnarray}
    E &=& \frac{1 - 2M/r_{\rm c}}{\sqrt{1 - 3M/r_{\rm c}}}\;,
    \nonumber\\
    L_z &=& \sqrt{\frac{r_{\rm c}M}{1 - 3M/r_{\rm c}}}\;,
    \nonumber\\
    \Gamma &=& \frac{r_{\rm c}^2}{\sqrt{1 - 3M/r_{\rm c}}}\;.
\end{eqnarray}
Combining these with Eq.\ (\ref{eq:quartic}) for $Q = 0$, $a = 0$, we find
\begin{eqnarray}
    \Omega_r &=& \sqrt{\frac{M}{r_{\rm c}^4}}\sqrt{r_{\rm c} - 6M}\;,
    \\
    T_r^{\rm GW} &=& \frac{5}{192{\dot{\cal E}}\eta M^3}\frac{r_{\rm c}^{9/2}(r_{\rm c} - 6M)^2}{\sqrt{r_{\rm c}-3M}(2r_{\rm c}^2 - 17Mr_{\rm c} + 42M^2)}\;.
    \nonumber\\
\end{eqnarray}
Their product is thus
\begin{equation}
    \Omega_r T_r^{\rm GW} = \frac{5r_{\rm c}^{5/2}}{192{\dot{\cal E}}\eta M^{5/2}}\frac{(r_{\rm c} - 6M)^{5/2}}{\sqrt{r_{\rm c} - 3M}(2r_{\rm c}^2 - 17Mr_{\rm c} + 42M^2)}\;.
\end{equation}
For circular Schwarzschild orbits, the transition is expected close to the ISCO at $r_{\rm c} = 6M$.  Putting $r_{\rm c} = (6 + x)M$ and expanding in $x$, we find
\begin{equation}
    \Omega_r T_r^{\rm GW} \simeq \frac{5x^{5/2}}{32\sqrt{2}\dot{\cal E}\eta}\;.
\end{equation}
Setting this equal to 1 and solving for $x$ yields
\begin{equation}
    x \simeq 4\frac{2^{1/5}{\dot{\cal E}}^{2/5}\eta^{2/5}}{5^{2/5}}\;. 
\end{equation}
This indicates that for quasi-circular inspiral, the transition begins at a distance from the ISCO that scales with mass ratio as $\eta^{2/5}$.  For the Schwarzschild value $\dot{\cal E} \simeq 1.14$, we find
\begin{equation}
    x \simeq 0.0639\left(\frac{\eta}{10^{-4}}\right)^{2/5}\;.
\end{equation}
This numerical value disagrees slightly with the result found in Appendix A of AH2019, with the difference arising from the incorrect conversion of time variables used there. (AH2019 also allow the transition to occur when $\Omega_r T_r^{\rm GW} \simeq A$, considering $A$ up to $\sim 10$.)

\subsection{Equation of motion for transition}
Once the system has entered the transition from inspiral to plunge, how do we describe the smaller body's motion?  Our ansatz is that the motion remains close to geodesic, but we allow the parameters of this geodesic to slowly evolve.  We begin with the radial geodesic equation
\begin{equation}
\left(\frac{dr}{d\lambda}\right)^2 = R(r)\;.
\end{equation}
In the absence of radiation reaction, this is zero for a circular orbit of radius $r = r_{\rm c}$.  We operate on both sides with $d/d\lambda$, yielding
\begin{eqnarray}
&&2\left(\frac{dr_{\rm c}}{d\lambda}\right)\left(\frac{d^2r_{\rm c}}{d\lambda^2}\right) =
\frac{\partial R}{\partial r}\left(\frac{dr_{\rm c}}{d\lambda}\right)
\nonumber\\
&&\qquad+ \frac{\partial R}{\partial E}\left(\frac{dE}{d\lambda}\right) + \frac{\partial R}{\partial L_z}\left(\frac{dL_z}{d\lambda}\right) + \frac{\partial R}{\partial Q}\left(\frac{dQ}{d\lambda}\right)\;.
\nonumber\\
\label{eq:circevol}
\end{eqnarray}
Let us carefully examine various terms which enter into Eq.\ (\ref{eq:circevol}):

\begin{itemize}

\item If the system were precisely geodesic, $\partial R/\partial r$ would vanish.  However, because the system is evolving through the transition, the integrals of motion are shifted away from geodesic values by an amount that scales with the mass ratio $\eta$.  As a consequence, $\partial R/\partial r$ is non-zero by a value of order $\eta$.

\item The GW-driven rates of change $dE/d\lambda$, $dL_z/d\lambda$, $dQ/d\lambda$, and $dr_{\rm c}/d\lambda$ are all of order $\eta$.  During inspiral, the GW-driven acceleration $d^2r_{\rm c}/d\lambda^2$ is of order $\eta^2$; during the transition this acceleration is of order $\eta$.

\item The derivatives $\partial R/\partial E$, $\partial R/\partial L_z$, and $\partial R/\partial Q$ are all independent of $\eta$.

\end{itemize}

\noindent
Combining these points, Eq.\ (\ref{eq:circevol}) breaks into terms that scale as $\eta$, and terms that scale as $\eta^2$:
\begin{eqnarray}
O(\eta):& & \frac{\partial R}{\partial E}\left(\frac{dE}{d\lambda}\right) + \frac{\partial R}{\partial L_z}\left(\frac{dL_z}{d\lambda}\right) + \frac{\partial R}{\partial Q}\left(\frac{dQ}{d\lambda}\right) = 0\;,
\nonumber\\
\label{eq:circconstraint}\\
O(\eta^2):& & \frac{d^2r_{\rm c}}{d\lambda^2} = \frac{1}{2} \frac{\partial R}{\partial r}\;.
\label{eq:circdynamics}
\end{eqnarray}
Equation (\ref{eq:circconstraint}) is a constraint that holds whenever we have adiabatically evolving circular orbits; Eq.\ (\ref{eq:circdynamics}) is a tool that allows us to characterize the radial behavior of the system as we begin evolving beyond the adiabatic regime.

Let us introduce an expansion of all physically relevant quantities as we approach the last stable orbit:
\begin{eqnarray}
    r_{\rm c} &=& r_{\rm ISCO} + x\;,
    \\
    E &=& E^{\rm ISCO} + \delta E\;,
    \\
    L_z &=& L_z^{\rm ISCO} + \delta L_z\;,
    \\
    Q &=& Q^{\rm ISCO} + \delta Q\;.
\end{eqnarray}
Here ``ISCO'' stands for innermost stable circular orbit, the last stable orbit for circular configurations.  Exactly at the ISCO, $R = \partial R/\partial r = \partial^2R/\partial r^2 = 0$.  Bearing this in mind, the leading order behavior of $R$ near the ISCO is given by
\begin{eqnarray}
R &\simeq& \frac{1}{6}\left(\frac{\partial^3 R}{\partial r^3}\right)x^3
\nonumber\\
&+& \left(\frac{\partial^2R}{\partial r\partial E}\delta E + \frac{\partial^2R}{\partial r\partial L_z}\delta L_z + \frac{\partial^2R}{\partial r\partial Q}\delta Q\right)x\;.
\nonumber\\
\end{eqnarray}
We leave off here terms that are independent of $x$, as well as terms higher order than those shown here.  All quantities in parentheses are to be evaluated exactly on the ISCO.

We imagine that the integrals of motion evolve linearly with time as we move through the transition, writing
\begin{eqnarray}
    \delta E &=& \eta\left(\frac{dE}{d\lambda}\right)^{\rm ISCO}(\lambda - \lambda_{\rm ISCO})\;,
    \nonumber\\
    \delta L_z &=& \eta\left(\frac{dL_z}{d\lambda}\right)^{\rm ISCO}(\lambda - \lambda_{\rm ISCO})\;,
    \nonumber\\
    \delta Q &=& \eta\left(\frac{dQ}{d\lambda}\right)^{\rm ISCO}(\lambda - \lambda_{\rm ISCO})\;.
\end{eqnarray}
This is essentially an assumption that the transition is short enough that a linear-in-time approximation effectively describes the integrals ($E$, $L_z$, $Q$) during this epoch of the binary's evolution.  We have introduced $\lambda_{\rm ISCO}$, the Mino time at which the secondary reaches the ISCO.  Notice that the rates of change are per unit $\lambda$.  For compactness, let us denote $\kappa_\mathcal{C} \equiv (d\mathcal{C}/d\lambda)^{\rm ISCO}$, where $\mathcal{C}$ can be $E$, $L_z$, or $Q$.  Evaluating the acceleration equation (\ref{eq:circdynamics}) under the assumption that all quantities are close to the ISCO yields the equation of motion
\begin{equation}
    \frac{d^2x}{d\lambda^2} = -A x^2 + \eta B(\lambda - \lambda_{\rm ISCO})\;,
    \label{eq:GOTstep1}
\end{equation}
where
\begin{eqnarray}
    A &\equiv& -\frac{1}{4}\left(\frac{\partial^3R}{\partial r^3}\right)\;,
    \label{eq:GOT_Adef}
    \\
    B &\equiv& -\frac{1}{2}\left(\frac{\partial^2R}{\partial r\partial E}\kappa_E + \frac{\partial^2R}{\partial r\partial L_z}\kappa_{L_z} + \frac{\partial^2R}{\partial r\partial Q}\kappa_Q\right)\;.
    \nonumber
    \label{eq:GOT_Bdef}\\
\end{eqnarray}
We next rescale the radial variable $x$ and the time variable $\lambda$ as
\begin{eqnarray}
    x/M &=& \eta^{2/5} B^{2/5}A^{-3/5} X\;,
    \\
    M(\lambda - \lambda_{\rm ISCO}) &=& \eta^{-1/5}(AB)^{-1/5}L\;.
\end{eqnarray}
With these scalings, we rewrite Eq.\ (\ref{eq:GOTstep1}) as
\begin{equation}
    \frac{d^2X}{dL^2} = -X^2 - L\;.
    \label{eq:ot_trans}
\end{equation}
When specialized to equatorial orbits, and working with proper time $\tau$ rather than Mino time, this is the transition equation derived by OT00; this generalization (using Mino time and not confining the analysis to equatorial orbits) was derived by AH19.  Equation (\ref{eq:ot_trans}) need only be solved once; the solution is discussed in Sec.\ III of AH19. The solution $X(L)$ then describes the transition for all quasi-circular inspirals to plunge (modulo the requirement that the mass ratio $\eta$ be small enough for perturbation theory to accurately describe the system).  The parameters pertaining to a particular system determine the constants $A$ and $B$ defined in Eqs.\ (\ref{eq:GOT_Adef}) and (\ref{eq:GOT_Bdef}).  Using $A$ and $B$, along with the mass ratio, it is simple to rescale to determine a particular system's $x(\lambda)$ for the transition, and from there to determine $r(t)$.  By connecting this transition trajectory $r(t)$ to a plunge, one can compute the complete worldline (as seen by distant observers) of an infalling body.  This procedure is employed in Refs. \cite{Lim2019, Hughes2019, Rifat2020, Islam2022} to develop small-mass-ratio inspiral and plunge worldlines, which in turn are used to produce inspiral-merger-ringdown waveforms in this limit.

\bibliographystyle{unsrt}

\bibliography{EccTrans}

\end{document}